\documentclass[
a4paper,
reprint,
showpacs,
showkeys,
 amsmath,amssymb,
 aps,
 prc,
floatfix,
]{revtex4-1}

\usepackage[utf8]{inputenc}
\usepackage{graphicx}
\usepackage{color}

\usepackage{qmech-revtex}

\begin{document}


\title{Energy, centrality and momentum dependence of dielectron
  production at collider energies in a coarse-grained transport
  approach}

\author{Stephan Endres}
 \email{endres@th.physik.uni-frankfurt.de}
\author{Hendrik van Hees}%
\author{Marcus Bleicher}%
\affiliation{%
Frankfurt Institute for Advanced Studies,
Ruth-Moufang-Straße 1, D-60438 Frankfurt, Germany
}%
\affiliation{
Institut f{\"u}r Theoretische Physik, Universit{\"a}t Frankfurt, Max-von-Laue-Straße 1, D-60438 Frankfurt, Germany
}

\date{\today}

\begin{abstract}
  Dilepton production in heavy-ion collisions at collider
  energies---i.e., for the Relativistic Heavy-Ion Collider (RHIC) and
  the Large Hadron Collider (LHC)---is studied within an approach that
  uses coarse-grained transport simulations to calculate thermal
  dilepton emission applying in-medium spectral functions from hadronic
  many-body theory and partonic production rates based on lattice
  calculations. The microscopic output from the Ultra-relativistic
  Quantum Molecular Dynamics (UrQMD) model is hereby put on a grid of
  space-time cells which allows to extract the local temperature and
  chemical potential in each cell via an equation of state. The
  resulting dilepton spectra are in good agreement with the experimental
  results for the range of RHIC energies,
  $\sqrt{s_{NN}}=19.6 - 200$\,GeV. The comparison of data and model
  outcome shows that the newest measurements by the PHENIX and STAR
  Collaborations are consistent and that the low-mass spectra can be
  described by a cocktail of hadronic decay contributions together
  with thermal emission from broadened vector-meson spectral functions
  and from the Quark-Gluon Plasma phase. Predictions for dilepton
  results at LHC energies show no significant change of the spectra as
  compared to RHIC, but a higher fraction of thermal contribution and
  harder slopes of the transverse momentum distributions due to the
  higher temperatures and flow obtained.
\end{abstract}

\pacs{25.75.Cj, 24.10.Lx}

\keywords{Dilepton production, Monte Carlo simulations}

\maketitle

\section{\label{sec:Intro}Introduction} 

A better understanding of the phase structure of strongly interacting
matter given by the fundamental theory of Quantum Chromodynamics (QCD)
is one of the main goals of heavy-ion experiments at ultra-relativistic
energies \cite{Stoecker:1986ci, Danielewicz:2002pu, Shuryak:2008eq,
  Rapp:2009yu, Schukraft2015CERN}. The collision of two nuclei produces
a fireball of hot and dense matter, which typically lives for a time
span of several $\fm/c$ until the system has cooled, due to collective
expansion, to a point where the single particles do not further interact
(freeze-out) \cite{Bjorken:1982qr}. The trajectory of the system within
the QCD phase diagram is determined by the collision energy: While for
lab-frame energies of few GeV one obtains rather low temperatures but
finds high values of baryochemical potential, the situation becomes
different when going to much higher collision energies; here the temperature
increases while the baryochemical potential decreases.

As hadronic observables usually only reflect the properties of the
system at the moment of freeze-out, electromagnetic probes are the
appropriate tool to obtain information from the earlier stages of the
reaction, when the system is at high temperatures and/or net-baryon
densities \cite{Feinberg:1976ua, Shuryak:1978ij}. Since photons and
dileptons do not interact strongly, they leave the fireball undisturbed
once they are produced. However, in consequence the measurement of
electromagnetic probes only gives a time integral over the various
stages and sources during the evolution of the reaction. From the
theoretical viewpoint the understanding of the production of
electromagnetic probes in a heavy-ion collision is complicated by the
fact that the evolving fireball of hot and dense matter is not a static
but a highly dynamical non-equilibrium system. However, no fully
self-consistent approach to describe the in-medium dilepton production
for the out-of-equilibrium case is available today. Consequently one has
to apply model descriptions, which always means a reduction of the
complexity of the problem to a level where it can be solved.

While hydrodynamic \cite{Vujanovic:2013jpa, Ryblewski:2015hea} and
fireball models \cite{Rapp:2013nxa, Rapp:2014hha} are in general successful in describing the measured dilepton spectra by the STAR
\cite{Adamczyk:2013caa,Adamczyk:2015lme,Adamczyk:2015mmx} and PHENIX
\cite{Adare:2015ila} collaborations at RHIC, these models completely
rely on a macroscopic description of the fireball. The application of
thermal emission rates is usually straightforward in these models, but
they require external assumptions such as an initial state and an
additional description for the final-state interactions. Besides, their
application at lower temperatures and densities is questionable. On the
other hand, a full microscopic description of the electromagnetic
emission---as it is realized in transport models \cite{Schmidt:2008hm,
  Weil:2012ji, Weil:2014rea, Endres:2013nfa, Bratkovskaya:2013vx} based
on kinetic theory---is theoretically challenging, especially at very
high collision energies. On the one hand, a fully coherent
implementation of the different interfering processes and a correct
off-shell treatment of the particles has not yet been obtained; on
the other hand, it is also still unknown how the microscopic transition
from the hadronic to the partonic phase (and vice versa) is actually
realized in QCD. Nevertheless there exist several approaches which aim
for such an advanced microscopic description including off-shell and
medium effects \cite{Schenke:2005ry, Schenke:2006uh, Weil:2012qh,
  Bratkovskaya:1997mp, Bratkovskaya:2007jk, Linnyk:2011vx,
  Linnyk:2015rco}.

The coarse-graining approach, which is used in the present work for the
theoretical calculation of dilepton production, is based on the concepts
presented in Ref.\ \cite{Huovinen:2002im} and has been successfully
applied to describe spectra of electromagnetic probes at SIS\,18, FAIR
and SPS energies \cite{Endres:2014zua, Endres:2015fna, Endres:2015egk,
  Steinheimer:2016vzu}. It offers a compromise between the microscopic
and macroscopic description of the collision evolution. On the one hand
the dynamics is here based on a purely microscopic description from the
Ultrarelativistic Quantum Molecular dynamics (UrQMD) model
\cite{Bass:1998ca, Bleicher:1999xi}, on the other hand the
``coarse-graining'' (i.e, the reduction of the large amount of
information regarding the phase-space coordinates of the single hadrons)---performed
by averaging over a large ensemble events and extracting the local
thermodynamic properties of the system---allows to describe the reaction
dynamics in macroscopic terms of temperature and chemical
potential. However, it has the advantage that it is in principle
applicable to all phases of a heavy-ion collision and also works for lower
collision energies where the use of other macroscopic models is
questionable.

For the present paper previous studies are extended to energies
available at the Relativistic Heavy-Ion Collider (RHIC) and the Large
Hadron Collider (LHC), which covers the range of center-of-momentum
energies from $\sqrt{s_{NN}}=19.6\,$GeV up to 5500$\,$GeV. In this
energy regime the net-baryon density is assumed to be close to zero for
the greatest part of the fireball evolution, and a significant amount of
the electromagnetic emission will stem from the Quark-Gluon Plasma
(QGP). The specific conditions found at these collision energies offer
the possibility to study---among others---the following issues:
\begin{itemize}
\item The experimental dilepton measurements will show whether the
  hadronic spectral functions, which have proven to successfully
  describe the low-mass dilepton excess, are also consistent with the
  conditions found in heavy-ion collisions at collider energies, where
  the baryochemical potential is significantly lower than the
  temperature for the greatest part of the reaction evolution. Previous
  work has shown that the in-medium effects on the spectral properties
  of baryon resonances should still play an important role since the
  modification of vector mesons is governed by the \textit{sum} of the
  baryon and anti-baryon densities, not the net density
  \cite{Rapp:2013nxa}.
\item At higher invariant masses ($M_{\text{e}^{+}\text{e}^{-}} >1 \,\GeV/c^{2}$)
  correlated open-charm decays give a significant contribution to the
  measured dilepton yield for RHIC and LHC energies
  \cite{Adare:2008ac}. Similar to the light vector mesons, whose
  spectral shape is modified in the medium, the charm contribution is
  known to be affected by the presence of a hot and dense medium
  \cite{vanHees:2004gq, Lang:2012cx, Lang:2013wya}. However, it is
  unclear how strong these effects are. A direct measurement is
  difficult, as one also finds a strong thermal contribution from the
  QGP in that mass region. It is therefore an important theoretical task
  to disentangle the different contributions and to provide a
  comprehensive description of the measured dilepton spectra. Although
  we do not consider charm contributions in the present study, the
  thermal results for the QGP contribution may serve as a baseline and
  help to limit the possible medium modifications for $D$ and $\bar{D}$
  mesons.
\item Due to the very high temperatures reached at the collider energies
  considered here, the partonic contribution to the overall dilepton
  yield will be much more dominant than at lower energies. This might
  facilitate to study the properties of the Quark-Gluon Plasma, e.g.,
  its temperature \cite{David:2006sr, Rapp:2013nxa}.
\item Further, it will be interesting whether the reaction dynamics of
  the colliding system shows deviations as compared to the situation at
  lower energies. Large parts of the evolution are dominated by the
  Quark-Gluon Plasma, in contrast to the situation at SPS or even
  SIS\,18 and FAIR. Experimental results for RHIC exhibited an
  unexpected large flow for direct photons, which is not fully
  explained by theory up to now \cite{Adare:2011zr, Adare:2015lcd}. 
  With regard to the coarse-graining approach it will be especially 
  interesting to see in how far the underlying microscopic dynamics, 
  which is completely hadronic, can account for the correct expansion 
  of the system and the time-evolution of temperature and chemical 
  potential.
\end{itemize}

The last aspect also points out a caveat. Whereas the creation of a
deconfined phase with free quarks and gluons is assumed to take place in
the early stages of heavy-ion collisions at RHIC and LHC energies, the
microscopic dynamics from UrQMD does not include a description of this
partonic phase. Nevertheless, we will argue that it is possible to
extract a reasonable and realistic picture of the fireball evolution and
thermal electromagnetic emission also at RHIC and LHC energies,
including a description of emission from a QGP phase. While a lattice
EoS and partonic rates can be applied to approximate the thermodynamic
properties and emission patterns inside a partonic phase, it is on the
other hand clear that the fireball evolution itself from the
coarse-grained dynamics can not reflect any effects due to the creation
of a Quark-Gluon Plasma on the microscopic level. Although it is assumed
that the influences of a phase-transition or crossover on the gross
microscopic evolution are not very significant, this is of course a
limiting factor of the present model. Nevertheless, the results might
help to understand if and how the phase structure of QCD is reflected in
the microscopic dynamics, if the comparison of the model outcome to
experimental data shows significant deviations.

This paper is structured as follows. In Sec.\,\ref{sec:Model} the
coarse-graining approach is introduced, and the various
dilepton-production mechanisms, which enter the calculations, are
outlined. This is followed by a presentation of the results for the
space-time evolution of the reaction (Sec.\,\ref{ssec:fireball}) and
dilepton spectra for RHIC and LHC energies (in \ref{ssec:RHICresults}
and \ref{ssec:LHC}). A comparison of the results for RHIC and LHC is
given in Sec.\,\ref{ssec:Comp}. Finally, we conclude with a summary and
an outlook to further studies in Sec.\,\ref{sec:Outlook}.

\section{\label{sec:Model} The Coarse-graining approach} 

In the following, the basic features of the coarse-graining approach are
outlined. This description is kept concise here, as the same model
was in detail presented previously; for details we refer the reader to
references \cite{Endres:2014zua, Endres:2015fna}.

\subsection{\label{ssec:UrQMD} Microscopic simulations} 

As a first step, simulations for the different collision energies are
conducted with the present version 3.4 of the Ultra-relativistic Quantum
Molecular Dynamics (UrQMD) approach \cite{Bass:1998ca, Bleicher:1999xi,
  Petersen:2008kb, UrQMDweb}, a semi-classical hadronic transport model
based on the principles of kinetic theory, in which the evolution of a
heavy-ion collision is described by the propagation of on-shell
particles on classical trajectories in combination with a probabilistic
treatment of the individual hadron-hadron scatterings. It constitutes an
effective solution of the Boltzmann equation, where the collision term
includes elastic and inelastic scatterings as well as resonance
decays. To account for quantum effects, the particles are represented by
Gaussian wave packets and effects such as Pauli blocking are
included. For hadron-hadron collisions with energies above
$\sqrt{s}=3$\,GeV the excitation of strings is possible. The model
includes all relevant meson and baryon resonances up to a mass of
$2.2\,\GeV/c^{2}$. Resonance parameters and cross-sections are adapted
and extrapolated to the values collected by the Particle Data Group
\cite{Yao:2006px}.

For being able to deduce a realistic fireball evolution in terms of $T$
and $\mu_{\mathrm{B}}$ and---in consequence---meaningful dilepton
spectra from the UrQMD simulation, one first has to check whether the
model can describe the bulk results measured in experiment. In general,
the UrQMD model has proven to describe the hadronic observables from
heavy-ion reactions very well in a wide range of collision
energies. Also up to RHIC and LHC energies the hadron production and the
resulting yields, ratios, rapidity and transverse-momentum spectra are
quite well described in the approach; for details we refer the reader to
references \cite{Bratkovskaya:2004kv, Petersen:2008kb,
  Mitrovski:2008hb}.  However, looking at specific observables one also
finds deviations of the model results from the experimental data.  This
is especially the case for the elliptic flow, $v_{2}$: Whereas the
elliptic flow is described quite well up to SPS energies, for higher
collision energies the average elliptic flow $\langle v_{2}\rangle$
underestimates the experimental results. At top RHIC energy of
$\sqrt{s_{NN}}=200\,\GeV$ the transport model reaches only roughly 60\%
of the measured value \cite{Petersen:2006vm}. Regarding the
transverse-momentum dependence of $v_{2}$, the under-prediction is most
prominent for high $p_{t}$ \cite{Bleicher:2000sx, Zhu:2005qa,
  Zhu:2006fb}. Nevertheless, the model reproduces the centrality
dependence and the gross features of the $v_{2}$ particle-type
dependence, such the mass-ordering for low $p_{t}$ and the
number-of-constituent-quark scaling for higher transverse momenta
\cite{Lu:2006qn}. Since the build-up of $v_{2}$ in the model correlates
to the rescattering rate, the low values of this observable in UrQMD can
be interpreted as a hint that a strongly interacting phase of partons is
created in the early reaction evolution \cite{Kolb:2000fha,
  Konchakovski:2012yg, Uphoff:2014cba} (see also
Sec.\,\ref{ssec:noneq}).

However, the anisotropic flow effects are very small (at the order of
few percent) and have only very little influence on the dilepton
invariant-mass and transverse-momentum spectra. In consequence, the
deviations from the experimental measurements will not play a
significant role for our present study. This is of course different for
studies of the anisotopic flow of electromagnetic probes, where the
deviations from the measured bulk $v_{2}$ will be apparent. To reproduce
these measurements, one will probably need an advanced description which
includes the effects of the partonic phase on the fireball evolution.

\subsection{\label{ssec:CGapproach} Extracting thermodynamic properties} 
Note that within the UrQMD model one has a well determined phase-space
distribution function $f(\vec{x},\vec{p},t)$, as the location and
momenta of all particles are known. However, since the full microscopic
treatment of the medium effects is quite complicated, the present
approach aims to reduce (i.e., to coarse-grain) the amount of
information given by $f(\vec{x},\vec{p},t)$, such that one can switch
from a microscopic to a macroscopic description of the
collision. Instead of the individual particle coordinates, the system is
then defined by its thermodynamic properties. To do so, it is first
necessary to obtain a smooth distribution function, which is realized by
averaging over a large number of events:
\begin{equation} f(\vec{x},\vec{p},t)=\left\langle \sum_{h}
\delta^{(3)}(\vec{x}-\vec{x}_{h}(t))
\delta^{(3)}(\vec{p}-\vec{p}_{h}(t))\right\rangle.
\label{eq:distfunc}
\end{equation} 
Here the angle brackets $\left\langle \cdot \right\rangle$ denote the
ensemble average. It is important to bear in mind that the UrQMD model
constitutes a non-equilibrium approach, whereas the thermodynamic
properties are well defined only for equilibrated matter. Consequently,
the approximate extraction of equilibrium quantities is consistent only
locally. Thus a grid of small space-time cells is set-up
where---following Eq.\,\ref{eq:distfunc}---for each of these cells the energy-momentum tensor and the baryon current are extracted as
\begin{equation}
\begin{split} T^{\mu\nu}&=\frac{1}{\Delta V}\left\langle
\sum\limits_{i=1}^{N_{h} \in \Delta 
p^{\nu}_{i}}{p^{0}_{i}}\right\rangle, \\
j^{\mu}_{\mathrm{B}}&=\frac{1}{\mathrm{\Delta} V}\left\langle
\sum\limits_{i=1}^{N_{\mathrm{B}/\bar{\mathrm{B}}} \in \Delta
V}\pm\frac{p^{\mu}_{i}}{p^{0}_{i}}\right\rangle.
\end{split}
\label{eq:barcurrtmunu}
\end{equation} 
Here $\Delta V$ is the volume of the cell, and the sum is taken over all (anti-)baryons or hadrons in the cell, respectively. If one knows $j^{\mu}_{\mathrm{B}}$ and
$T^{\mu\nu}$, the local rest frame (LRF) can be determined by applying
the definition of Eckart \cite{Eckart:1940te}, which requires a vanishing
baryon flow, $\vec{j}_{\mathrm{B}}=0$. The energy and net-baryon density
of the cell are then defined as $\varepsilon=T^{00}_{\mathrm{LRF}}$ and
the baryon density is $\rho_{\mathrm{B}}=j^{0}_{\mathrm{B, LRF}}$.

To obtain temperature and baryochemical potential it is necessary to
apply an equation of state (EoS) which translates the local energy and
baryon densities into $T$ and $\mu_{\mathrm{B}}$. For consistency with
the underlying transport model, we apply a hadron gas EoS
\cite{Zschiesche:2002zr} for the lower temperature range up to
$T = 170$\,MeV. It includes the same hadronic degrees of freedom as the
UrQMD approach. For higher temperatures a pure hadronic description is
insufficient, as the phase transition to a Quark-Gluon Plasma also
changes the degrees of freedom and consequently the equation of
state. We therefore use an EoS from lattice calculations
\cite{He:2011zx} (with a critical temperature $T_{c}=170$\,MeV) for
cells with higher energy densities. While both EoS match in the
temperature region from 150-170 MeV, the lattice EoS gives significantly
higher temperatures for very hot cells. A comparison between both
equations of state is given in Ref.\,\cite{Endres:2014zua}.

It is important to bear in mind that the application of the lattice EoS
for higher energy densities or temperatures, respectively, does not provide
full consistency with the underlying hadronic dynamics in the transport
model; in UrQMD only hadronic degrees of freedom are implemented, and no
phase transition to a partonic phase is included. On the other hand, the
very details of the microscopic dynamics are anyway ``washed out'' in
the coarse-graining procedure by the reduction of the multitude of information and
the averaging over the events. Since we only use the local energy
density distribution from the microscopic simulations to calculate a
temperature via the lattice EoS ($\mu_{\mathrm{B}}=\mu_{\pi}=0$ is
always assumed for $T>170$\,MeV), a severe problem should only arise if
the gross evolution of the density distribution would largely depend on
the specific equation of state. This would imply differences in the
measurable particle spectra. However, previous studies with a
UrQMD+hydrodynamics hybrid model \cite{Petersen:2008dd} have shown that
the bulk evolution of the fireball is not significantly altered when
using an EoS including a phase transition instead of a pure hadron gas
EoS \cite{Steinheimer:2009nn}. Taking this into account, the procedure
as applied in the present approach seems justifiable. (The effect of the
choice of EoS on the dilepton spectra is also studied in
Sec.\,\ref{ssec:RHICresults}.)

\subsection{\label{ssec:noneq} Non-equilibrium effects} 

The approach as outlined above assumes a locally equilibrated system in
each cell. However, it is clear that within a transport approach this
condition is not always fulfilled in a satisfying manner. In contrast,
due to the non-equilibrium nature of the model one finds significant
deviations from kinetic and/or chemical equilibrium. For a correct
description of the fireball evolution the consequences of these
deviations need to be considered. Basically one finds two dominant
effects which affect the thermodynamic properties and, consequently, the
dilepton emission:
\begin{enumerate}
\item Pressure isotropy is necessary for a system to be in
  \textit{kinetic equilibrium}. However, it is well known from previous
  studies \cite{Bravina:1999kd, Bravina:1999dh} that the initial stages
  of a heavy-ion collision are dominated by large differences between
  the longitudinal and transverse pressures. This is a consequence of
  the strong longitudinal compression of the nuclei at the beginning of
  the collision. In this case, the energy density is overestimated in
  the cell, as a large fraction of the energy is of no relevance with
  regard to the thermal properties of the system. To apply the
  coarse-graining approach also for the first few $\fm/c$ of the
  collision, it is therefore necessary to extract a realistic energy
  density $\varepsilon_{\mathrm{eff}}$ taking the limited degree of
  thermalization into account. This is achieved by the use of a
  generalized equation of state for a Boltzmann-like system
  \cite{Florkowski:2010cf, Florkowski:2012pf}, that gives
  $\varepsilon_{\mathrm{eff}}$ in dependence on the ``bare'' energy
  density in the cell and the pressure anisotropy. The results for SPS
  energies showed that significant deviations of
  $\varepsilon_{\mathrm{eff}}$ are only found for the first 1-2\,fm/$c$
  of the collision \cite{Endres:2014zua}.

\item \textit{Chemical non-equilibrium} shows up in the form of finite
  meson chemical potentials (in full equilibrium, all meson chemical
  potentials vanish as the meson number is not a conserved quantity, in
  contrast to, e.g., the net-baryon number) and most dominantly in form
  of a pion chemical potential $\mu_{\pi}$, since the $\pi$ mesons are
  the most abundantly produced particles. A finite $\mu_{\pi}$ is the
  consequence of an overpopulation of pion states. In a transport model,
  such an over-dense pion system is especially found at the very
  beginning of the reaction, when the fireball is still far from kinetic
  equilibrium and the first inelastic collisions produce a large number
  of pions \cite{Bandyopadhyay:1993qj}. The pion chemical potential is
  important for the population of $\rho$ and $\omega$ vector mesons, as
  a high density of pions increases the probability for the production
  of these particles (besides, $\mu_{\pi}$ has also some moderate
  effects on the spectral shape) \cite{Koch:1992rs, Rapp:1999ej}. To
  account for these effects we extract the pion chemical potential in
  each cell in Boltzmann approximation.
\end{enumerate}

When the local energy and particle densities change in the course of the
fireball evolution, the phase-space distribution function,
$f(\vec{x},\vec{p},t)$, is adjusted to the corresponding values of
temperature and chemical potentials. If this adjustment is slower than
the change of $T$ and $\mu$, one will find deviations from the local
equilibrium distribution of the form \cite{Heinz:2009xj}
\begin{equation}
f(\vec{x},\vec{p},t)=f_{\mathrm{eq}}(\vec{x},\vec{p},t)+\delta f (\vec{x},\vec{p},t).
\end{equation}
One consequence of this deviation from the equilibrium state is the
appearance of finite transport coefficients, such as viscous stresses,
heat flow, and diffusion \cite{Muronga:2003ta}. Note that in the
underlying transport dynamics used for the coarse-graining these effects
are implicitly implemented due to the non-zero mean free path of the
interacting hadrons. The resulting transport coefficients (e.g.,
viscosity and heat conductivity) from UrQMD have been in detail studied
for the infinite-matter case in box calculations \cite{Muronga:2003tb,
  Demir:2008zz, Demir:2008tr, Nemakhavhani:2016aje}. The results showed
that for the shear viscosity to entropy ratio one obtains rather high
values $\eta/s > 0.6$ within the model. This is in contrast to ideal
hydrodynamic calculations which have been quite successful in describing
the observables from heavy-ion collisions by neglecting the effect of
those transport coefficients. The large elliptic flow measured in
non-central heavy-ion reactions at RHIC energies suggests a very low
value of the shear viscosity to entropy ratio $\eta/s$ in the created
hot and dense fireball. This was interpreted as a direct hint for the
creation of a Quark-Gluon Plasma phase early during the fireball
evolution \cite{Heinz:2001xi, Gyulassy:2004zy}. The high values of
$\eta/s$ from UrQMD can, in consequence, explain the underestimation of
the resulting elliptic flow $v_{2}$ at RHIC in the model, as discussed
in Sec.\,\ref{ssec:UrQMD}.

However, recently the role and importance of viscosity has come into
theoretical focus and was studied intensively in hydrodynamical
approaches \cite{Teaney:2003kp, Dusling:2009bc, Dion:2011pp,
  Schenke:2006yp, Liu:2015mgo, Liu:2015vst}, as it was found that pure
ideal hydrodynamic calculations result in an overestimate of the
elliptic flow for high transverse momenta and/or wrong slopes for the
hadron-$p_{t}$ spectra \cite{Heinz:2009xj}. With regard to the dilepton
emission, the appearance of a shear viscosity might show an effect in
two ways: Firstly, by its influence on the bulk evolution---especially
an increase of the directed flow and a reduction of the anisotropies---,
and secondly by the direct modification of the emission rates due to
modifications of the distribution functions \cite{Shen:2013cca}.

Whereas the coarse-grained dynamics of the fireball naturally reflects
the viscosities in the underlying microscopic simulations, as mentioned
above, we do not consider the effects of the viscous corrections on the
electromagnetic emission rates for two reasons: On the one hand it was
shown that the influence of finite viscosity on the resulting invariant
mass or transverse momentum spectra of dileptons and photons is rather
small, especially for the low-mass region up to $2\,\GeV/c^{2}$
\cite{Ryblewski:2015hea, Vujanovic:2013jpa, Shen:2014nfa}. (The case is
somewhat different for the elliptic flow, where the modification of the
emission rates might be more pronounced.) On the other hand there are
presently no calculations available for the hadronic and partonic rates
which are applied in our approach. The viscous correction for emission
from the Quark-Gluon Plasma has so far only been calculated for the
perturbative Born rate, i.e., for leading order $q\bar{q}$ annihilation
\cite{Cleymans:1986na}. However, this rate is known to significantly
underestimate the thermal yield for lower masses, compared to more
advanced hard-thermal loop or lattice rates \cite{Braaten:1990wp,
  Ding:2010ga}. The situation is similar for the hadronic rates, where
the effect of viscosity has been considered only for a low-density
calculation \cite{Eletsky:2001bb} which cannot account for the full
in-medium modfications of the vector mesons' spectral shape.

We will discuss the emission rates applied in the present approach in
detail in the following section \ref{ssec:dilrates}.

\subsection{\label{ssec:dilrates}Thermal dilepton rates} 

The thermal emission of dileptons from an equilibrated system of hot and
dense matter is determined by the imaginary part of the (retarded)
electromagnetic current-current correlation function,
$\im \Pi_{\mathrm{em}}^{(ret)}$, which is connected to the electromagnetic
current $j_{\mu}$ \cite{Klingl:1997kf}. The dilepton yield per
four-volume and four-momentum can then be calculated according to the
relation \cite{McLerran:1984ay, Rapp:1999ej}
\begin{equation}
\label{rate_dil} \frac{\mathrm{d} N_{ll}}{\mathrm{d}^4x\mathrm{d}^4q} =
-\frac{\alpha_\mathrm{em}^2 L(M)}{\pi^3 M^2} \; f_{\mathrm{B}}(q;T) \im
\Pi^{(\text{ret})}_\mathrm{em}(M, \vec{q};\mu_B,T),
\end{equation} 
where $f_{\mathrm{B}}$ is the Bose distribution function and $L(M)$ the
lepton phase space.

In the hadronic low-mass regime (i.e., for $M_{\mathrm{e^{+}e^{-}}} < 1 \,\GeV/c^{2}$) the
electromagnetic current directly couples to the vector mesons
and---assuming vector meson dominance (VMD)---$\Pi_{\mathrm{em}}$ is
proportional to the vector-meson propagator
\begin{equation}
\label{vacprop} D_{V}=\frac{1}{q^{2}-m_{V}^{2}-\Sigma_{V}(q^{2})}
\end{equation} 
where $m_{V}$ is the bare mass of the meson and $\Sigma_{V}$ the
corresponding self-energy of the particle, related to its decay
width. Whereas the self-energy in the vacuum can be deduced from
experimental measurements of inelastic electron-positron scattering
($\mathrm{e}^{+}\mathrm{e}^{-} \rightarrow \mathrm{hadrons}$), the
situation for finite $T$ and $\mu_{\mathrm{B}}$ is more complicated and
requires detailed model calculations. For the present work we apply the
results from equilibrium quantum-field theory calculations with a
hadronic many-body approach \cite{Rapp:1999us, RappSF}. They account for
the interactions of the $\rho$ and $\omega$ mesons with hadrons in a
heat bath. For the $\rho$ the pion cloud ($\Sigma_{\rho\pi\pi}$) as well
as the direct contributions from $\rho$-hadron scatterings with baryons
($\Sigma_{\rho B}$) and mesons ($\Sigma_{\rho M}$) are included in the
calculation of the in-medium self-energy. In this case
Eq.\,\ref{vacprop} becomes
\begin{equation}
\label{rhoprop}
D_{\rho}=\frac{1}{M^{2}-m_{\rho}^{2}-\Sigma_{\rho\pi\pi}-\Sigma_{\rho
B}-\Sigma_{\rho M}}.
\end{equation} 
The situation for the $\omega$ meson is more complex, as it constitutes
a three-pion resonance. Here the self energy includes
$\omega \rightarrow \pi\rho$ and $\omega \rightarrow 3\pi$ decays as
well as the inelastic $\omega\pi \rightarrow \pi\pi$,
$\omega\pi \rightarrow b_{1}$ and $\omega N \rightarrow N^{*}$
scatterings. The resulting propagator reads
\begin{equation}
\begin{split}
\label{omprop} D_{\omega}=& [M^{2}-m_{\omega}^{2}+\mathrm{i} m_{\omega}\left(
\Gamma_{3\pi}+\Gamma_{\rho\pi}+\Gamma_{\omega\pi \rightarrow
\pi\pi}\right) \\& - \Sigma_{\omega\pi b_{1}} - \Sigma_{\omega B} ]^{-1}.
\end{split}
\end{equation} 
To account for the symmetry of the interactions of $\rho$ and $\omega$
mesons with baryons and anti-baryons, the spectral functions do not
depend on the baryochemical potential $\mu_{B}$ but on an effective
baryon density
$\rho_{\mathrm{B}}^{\mathrm{eff}} = \rho_{\mathrm{N}} +
\rho_{\bar{\mathrm{N}}} + 0.5 (\rho_{\mathrm{B^{*}}} +
\rho_{\bar{\mathrm{B}^{*}}})$ \cite{vanHees:2007th}. Here
$\rho_{\mathrm{N}/\bar{\mathrm{N}}}$ denotes the nucleon / anti-nucleon
density and $\rho_{\mathrm{B^{*}}/\bar{\mathrm{B}}^{*}}$ is the density
of excited baryon/anti-baryon resonances.

Note that in the case of a finite pion chemical potential an additional
fugacity factor
\begin{equation}
\label{fugacity} z^{n}_{\pi}=\exp\left(\frac{n\mu_{\pi}}{T}\right)
\end{equation} 
enters in Eq.\,\ref{rate_dil}. The exponent $n$ depends on the
difference between initial and final pion number for the relevant
channel \cite{Baier:1997td,Baier:1997xc,Rapp:1999ej}. For dilepton
production from $\rho$ mesons one has $n=2$ whereas for the $\omega$ it
is $n=3$.

At the higher masses above $1\,\GeV/c^{2}$ one no longer finds distinct
resonances in the hadronic domain of the vector channel but a broad 
continuum of multi-pion states which couple to the electromagnetic current. In principle, also
here the dilepton emission is related to the vector spectral
function. However, the presence of pions at finite $T$ causes a chiral
mixing of the isovector part of the vector and axial-vector correlators
\cite{Dey:1990ba}. The corresponding isovector-vector current
correlation function takes the form \cite{vanHees:2006ng}
\begin{equation}
\begin{split}
  \label{vamix} \Pi_{V}(p) = &
(1-\varepsilon)z_{\pi}^{4}\Pi^{\text{vac}}_{\text{V},4\pi} +
\frac{\varepsilon}{2}z^{3}_{\pi} \Pi^{\text{vac}}_{\text{A},3\pi} \\ &+
\frac{\varepsilon}{2}(z^{4}_{\pi}+z^{5}_{\pi})\Pi^{\text{vac}}_{\text{A},5\pi},
\end{split}
\end{equation} 
where the mixing coefficient $\varepsilon$ is given by the thermal pion
loop, and $z_{\pi}$ again denotes the pion fugacity.

For temperatures above the critical temperature $T_{c}$ the relevant
degrees of freedom are no longer hadrons (vector mesons) but quarks and
gluons. In this situation the strength of the electromagnetic current is
accounted for by a partonic description and the thermal dilepton
production occurs---to leading order---via the electromagnetic
annihilation of quark-antiquark pairs,
$q\bar{q} \rightarrow \gamma^{*}$. However, it has been shown that the
pure pQCD result \cite{Cleymans:1986na} underestimates the actual
dilepton emission in the low energy regime (i.e., at low
masses). Nonperturbative results indicate a strong enhancement due to
$\alpha_{s}$ corrections and bremsstrahlung effects
\cite{Braaten:1990wp}. In the present work we apply a spectral function
from lattice QCD calculations \cite{Ding:2010ga} which has been
extrapolated for finite three-momenta by a fit to the according photon
rate \cite{Rapp:2013nxa}. Note that these lattice rates are available only 
for vanishing quark chemical potential $\mu_{q}=0$. However, the effects of
a finite $\mu_{q}$ are quite small with regard to the dilepton emission rates
and can be neglected here.

\subsection{\label{ssec:decay} Non-thermal hadronic decay contributions}

In addition to the thermal dilepton emission from the hot and dense
fireball, there are also contributions from more long-lived mesons which
mostly decay into lepton pairs after the freeze-out of the system,
mainly the pseudoscalar $\pi^{0}$ and $\eta$ mesons. Their Dalitz decays
into a real and a virtual photon (which subsequently transforms in a
lepton pair) dominate the very low invariant masses. The corresponding
decay width is related to the probability for the decay into two photons
and given by the Kroll-Wada formula \cite{Kroll:1955zu}
\begin{equation}
\begin {split}
\label{krollwada} \frac{\dd \Gamma_{P\rightarrow \gamma e^{+}e^{-}}}{\dd
M} = & \frac{2\alpha}{3\pi M}\ L(M)\ 2\Gamma_{P\rightarrow \gamma\gamma}
\\ & \times\left(1-\frac {M^{2}}{M^{2}_{\rho}}\right) \vert
F_{P\gamma\gamma^{*}} (M^{2})\vert,
\end{split}
\end{equation} 
where the form factors $F_{P\gamma\gamma^{*}}$ are fitted to
experimental data \cite{Landsberg:1986fd}, consistent with the
theoretical results assuming VMD.
\begin{figure*}
\includegraphics[width=1.0\columnwidth]{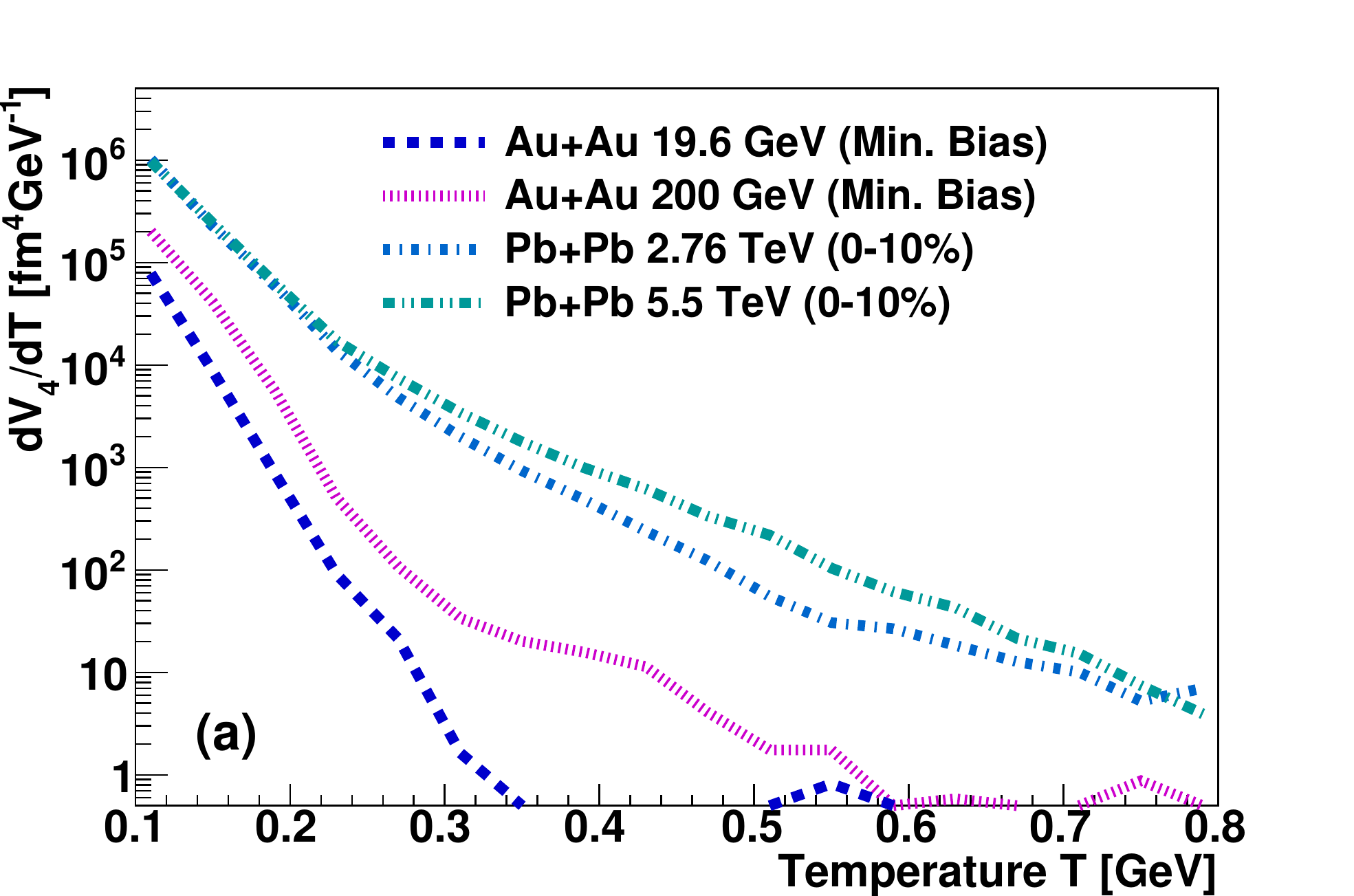}
\includegraphics[width=1.0\columnwidth]{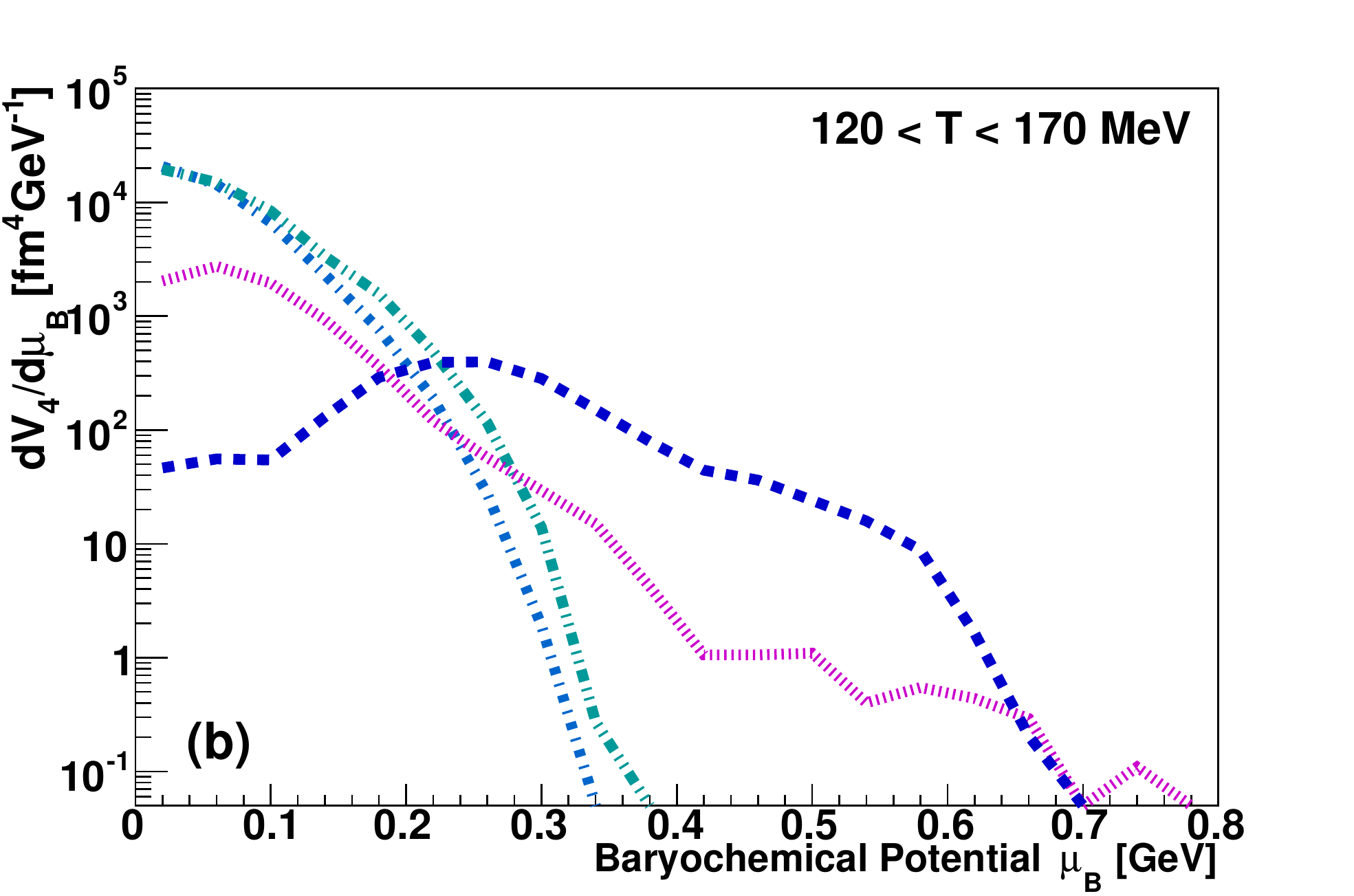}
\caption{(Color online) Thermal four-volume $V_{4}$ in dependence on
  temperature (a) and baryochemical potential (b) for Au+Au and Pb+Pb
  reactions at different collision energies.}
\label{thermfourvol}
\end{figure*} 
\begin{figure}[b]
\includegraphics[width=1.0\columnwidth]{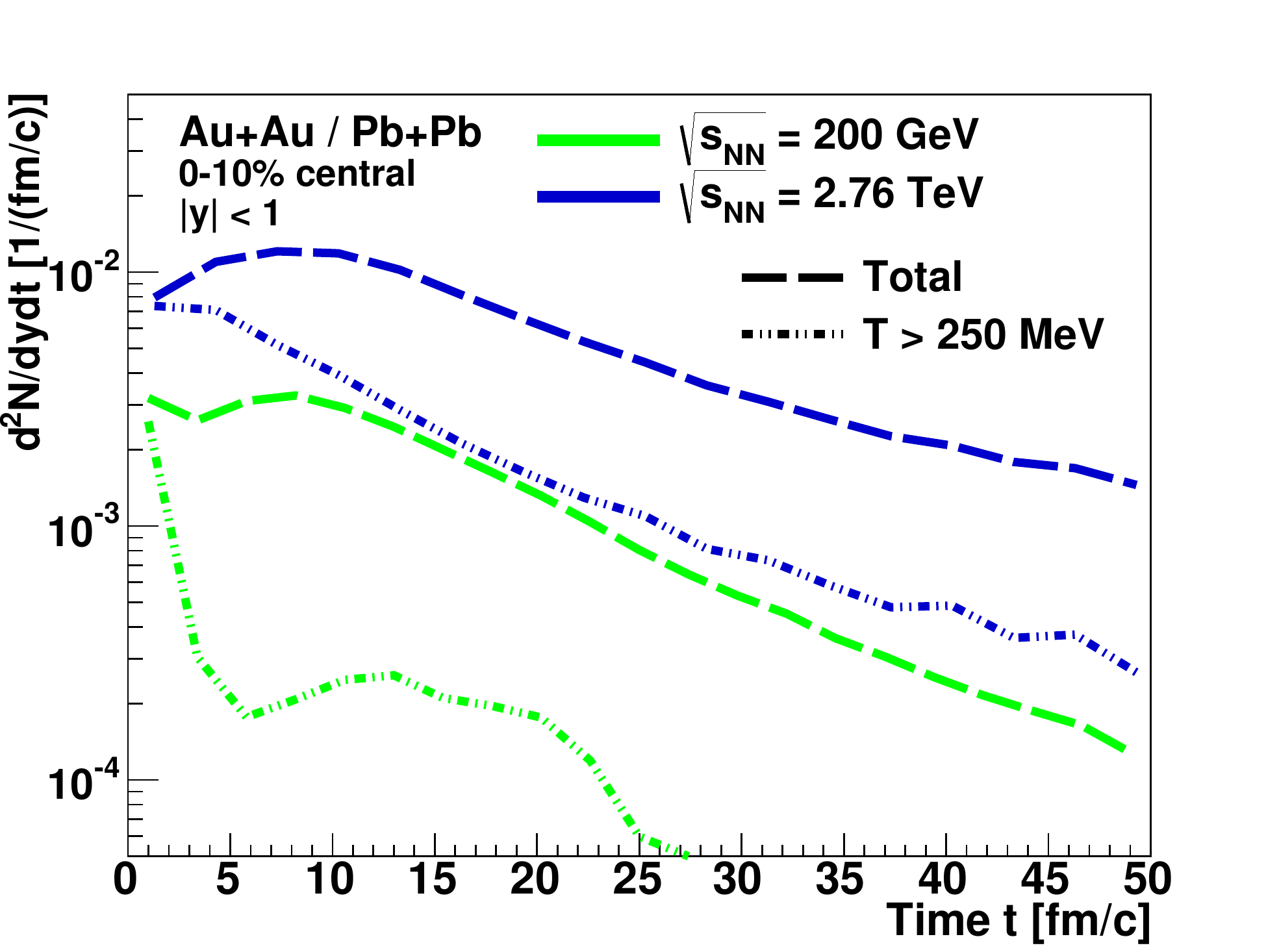}
\caption{(Color online) Time evolution of the thermal dilepton emission
  $\dd N_{\mathrm{e}^{+}\mathrm{e}^{-}}/\dd t$ for central Au+Au collisions at
  $\sqrt{s_{NN}}=200$\,GeV (green) and Pb+Pb collisions at
  $\sqrt{s_{NN}}=2.76$\,TeV (blue). The total emission (long dashed) is
  shown as well as the resulting yield only from cells with a
  temperature above 250\,MeV (dashed double-dotted).}
\label{fourvoltime}
\end{figure} 

Note that only the final state $\pi$ and $\eta$ mesons are considered
for the procedure. Those mesons which are produced and absorbed again
during the collision have a negligible probability for a dilepton decay
due to their small decay width. The situation is somewhat different for
the $\phi$ meson. In spite of the shorter lifetime we do not treat it as
a thermal contribution (since the expected medium-effects are so small
that they can be neglected) but consider the microscopic decays here as
for the pseudoscalar mesons. However, in this case one assumes that the $\phi$ has an equal
probability for the decay into a lepton pair at any time and therefore
can continuously emit dileptons \cite{Li:1994cj}. The total yield is
then obtained as a time integral over the lifetime as
\begin{equation}
\label{phishining} \frac{\mathrm{d}N_{ll}}{\mathrm{d}M} =\frac{\Delta
N_{ll}}{\Delta M}= \sum _{i=1}^{N _{\Delta M}}\sum _{j=1}^{N _{\phi}}
\int_{t_{i}}^{t_{f}} \frac{\mathrm{d}t}{\gamma}
\frac{\Gamma_{\phi\rightarrow ll}(M)}{\Delta M},
\end{equation} 
where the $\gamma$ factor accounts for the relativistic time dilation in
the computational frame compared to the mesons rest frame. This
procedure explicitly takes absorption processes for the $\phi$ into
account.

Besides, two more non-thermal contributions arise due to the fact that
not for all cells it is possible to properly calculate the thermal
contribution. This is mainly the case for the later stages of the
reaction, for cells with (i) no baryon content, so that the LRF is not
well-defined, or (ii) where the temperature is below 50 MeV, in which
case the EoS and the emission rates no longer give reliable results. In
these cases a ``freeze-out'' contribution for the $\rho$ and $\omega$
meson is determined directly from the microscopic UrQMD results for
those specific cells. The procedure is the same as for the $\phi$ given
by Eq.\,\ref{phishining}, but the time-integration is performed only for
the corresponding time-step size.  

\section{\label{sec:Results}Results} 
For the present study the coarse-graining of the UrQMD transport output
was performed with ensembles of 1000 UrQMD events for Au+Au collisions
at RHIC and 500 events for Pb+Pb reactions at LHC energies. The
time-step size was chosen as $\Delta t=0.4-0.6$\,fm/$c$, and the spatial
dimensions of the cell are defined as
$\Delta x = \Delta y = \Delta z = 0.8-0.9$\,fm, depending on the
collision energy. The impact parameter distributions corresponding to
different centrality classes were chosen using Glauber-Model fits to
experimental data \cite{Adamczyk:2015lme,Adare:2015ila}. Note that the
minimum bias definitions slightly differ between the STAR and PHENIX
collaborations; the former uses 0-80\% most central collisions whereas
the PHENIX trigger takes 0-92\% central collisions into account.

\subsection{\label{ssec:fireball}Fireball evolution} 
\begin{figure*}
\includegraphics[width=1.0\columnwidth]{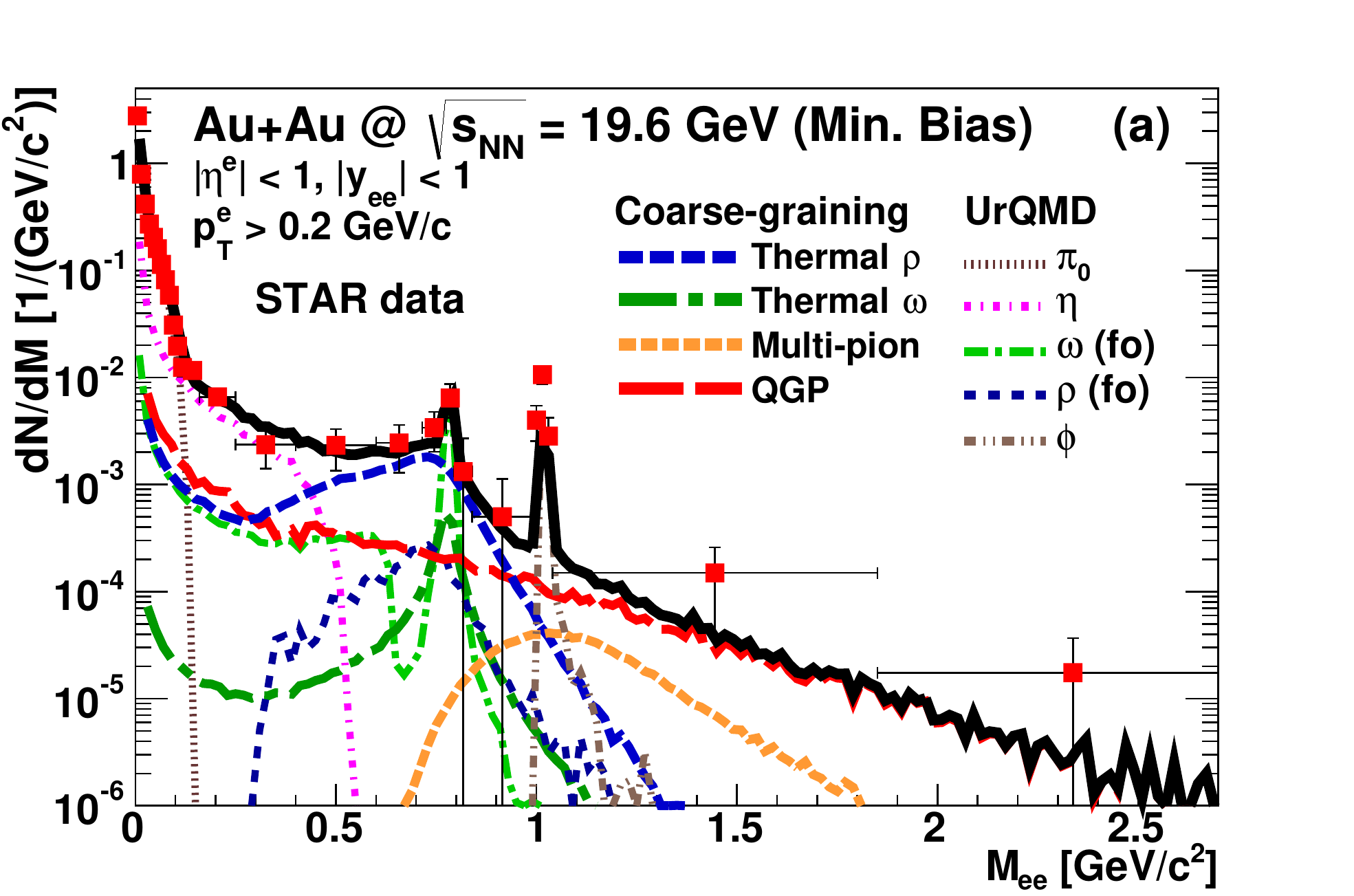}
\includegraphics[width=1.0\columnwidth]{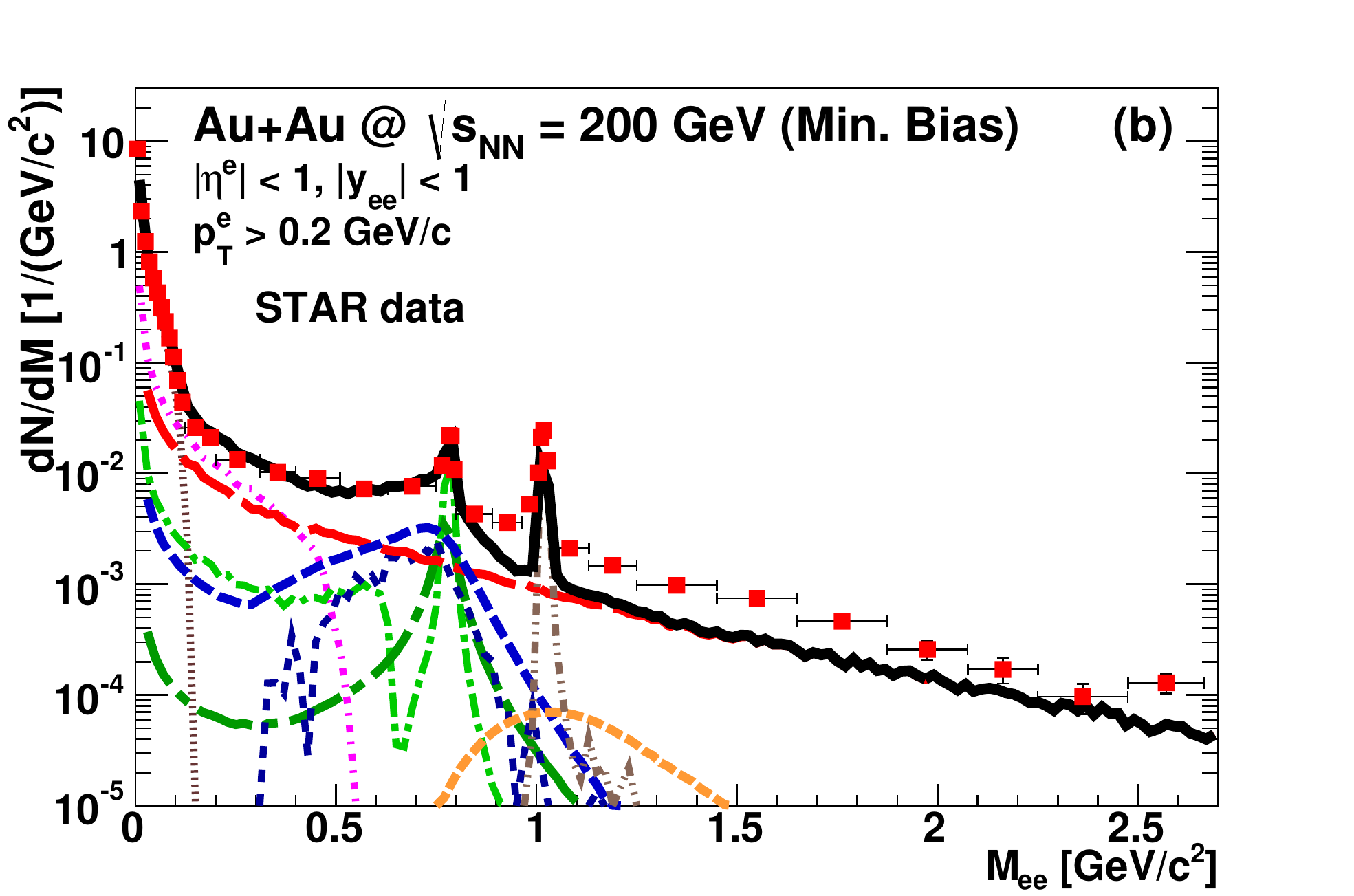}
\caption{(Color online) Dielectron invariant-mass spectra for minimum bias
  (i.e., 0-80\% most central) Au+Au collisions at $\sqrt{s_{NN}}=19.6$\,GeV
  (a) and $200\,$GeV (b). The sum includes the thermal hadronic and
  partonic emission obtained with the coarse-graining, and also the
  hadronic $\pi,\eta\text{ and }\phi$ decay contributions from UrQMD as
  well as the ``freeze-out'' contributions (from cold cells) of the
  $\rho$ and $\omega$ mesons. The model results are compared to the
  experimental data obtained by the STAR Collaboration
  \cite{Adamczyk:2015lme}.}
\label{dilinvmassRHIC}
\end{figure*}

The thermal dilepton emission from a hot and dense fireball created in a
heavy-ion collision is determined by the trajectory of the system within
the QCD phase diagram. More precisely, since for each space-time cell
different values of temperature and baryochemical potential are obtained
within the coarse-graining approach, the overall yield is directly
related to the distribution of the thermal four-volume $V_{4}$ inside
the fireball with regard to $T$ and
$\mu_{\mathrm{B}}$. Figure~\ref{thermfourvol}\,(a) shows the total
thermal four-volume summed over all cells in dependence on the
respective temperature for Au+Au and Pb+Pb reactions at four different
collision energies, from the lowest RHIC to top LHC energies. While for
the low temperature range around 100 MeV the differences between the
energies are not larger than one order of magnitude, the relative
increase of the number of higher temperature cells is much stronger. For
$\sqrt{s_{NN}}=19.6$\,GeV one hardly finds cells with temperature above
300\,MeV, while at LHC energies there are some cells with up to 800\,MeV
(few rare cells even reach still higher temperatures up to 1000\,MeV,
which is not shown here).

When considering the $\mu_{\mathrm{B}}$ dependence of the four-volume for the
temperature range from 120 to 170\,MeV in
Figure\,\ref{thermfourvol}\,(b), one also finds that the average baryon
chemical potential is decreasing when going to higher collision
energies (note again, as outlined in Sec.\,\ref{ssec:CGapproach}, the lattice EoS for $T>170$\,MeV in general assumes vanishing baryochemical potential). At $\sqrt{s_{NN}}=19.6$\,GeV the most abundant
$\mu_{\mathrm{B}}$-range lies between 200 and 300\,MeV, whereas at LHC
$\mu_{\mathrm{B}}$ is close to zero for the overwhelming part of the
thermal four-volume. Interesting is the fact that one gets a slightly
stronger contribution from higher chemical potential when going from
2.76 to 5.5\,GeV. However, this might be an effect due to the limited
temperature window considered here.

The resulting time evolution of the thermal dilepton emission
$\dd N/\dd t$ from all cells (and from those with temperature above
250\,MeV only) is shown in Figure\,\ref{fourvoltime}. The
results for central (0-10\%) Au+Au reactions at 200\,GeV and Pb+Pb
collisions at 2.76\,TeV exemplarily expose the similarities and differences in the fireball dynamics for RHIC and LHC. In general, one observes that the evolution of the
fireball for both energies is very similar, apart from the larger
overall emission at 2.76\,TeV compared to the 200\,GeV case. This is
a consequence of the larger thermal four-volume for all
temperature regions, compare Fig.\,\ref{thermfourvol}\,(a). However, at
the LHC the cooling of the system is slower, especially the emission
from the very hot cells with $T>250$\,MeV shows a less significant drop
than for the RHIC energy. In any case, the thermal emission from the
later stages of the reaction---even 40-50\,$\fm/c$ after the first
initial nucleon-nucleon interactions---is remarkably large, although the
influence on the total yield is very small, as $\dd N/\dd t$ is
suppressed by 1-2 orders of magnitude compared to the early maxima.

\subsection{\label{ssec:RHICresults} Relativistic Heavy-Ion Collider
(RHIC)}

The dilepton invariant-mass spectra for minimum bias Au+Au reactions at
the two RHIC energies $\sqrt{s_{NN}}=19.6 \text{ and } 200$\,GeV are
presented in Figure~\ref{dilinvmassRHIC}. The results as obtained with
the coarse-graining approach are compared to the experimental data from
the STAR Collaboration \cite{Adamczyk:2015lme}. The spectra are shown within the STAR
acceptance, which means rapidity and pseudorapidity cuts
($|\eta^{e}|<1$, $|y^{\text{ee}}|<1$) were applied for single electrons
and dileptons, respectively, together with an additional transverse
momentum cut for electrons (i.e., here $p_{t}^{e}>0.2$\,GeV). The
comparison shows that in both cases the invariant-mass spectra for low
masses below $1\,\GeV/c^{2}$ are very well described within the
model. While in comparison to pure hadronic decay cocktails an excess of
the experimentally measured spectra was observed for the mass region
$0.3 < M_{\text{ee}} < 0.7 \,\GeV/c^{2}$, our approach shows that this
region is dominated by thermal emission from the $\rho$ meson and from
partonic emission. But there are also important differences visible
when comparing the outcome for both energies: Due to the larger
temperatures obtained for Au+Au reactions at 200\,GeV, the low mass
region is here dominated by QGP emission, only around the $\rho$ pole
mass the hadronic emission is dominant. In contrast, the thermal $\rho$
contribution clearly outshines the partonic yield for the greatest part
of the low mass region up to 1\,GeV/$c^{2}$ at the lower collision
energy of 19.6\,GeV. 
\begin{figure}
\includegraphics[width=1.0\columnwidth]{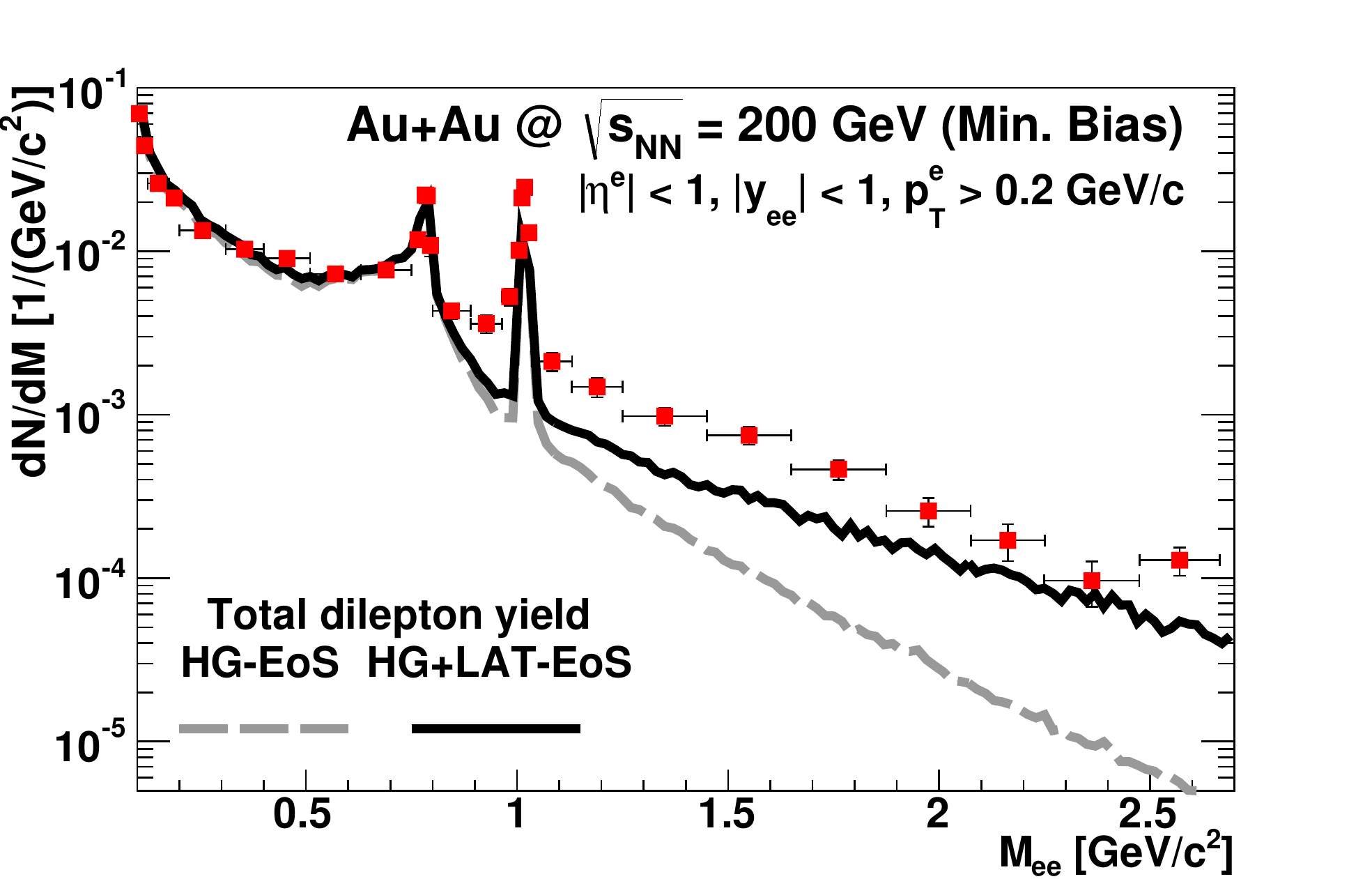}
\caption{(Color online) Comparison of the influence of different EoS on
  the dielectron invariant-mass spectrum for minimum bias Au+Au
  collsisions at $\sqrt{s_{NN}}=200\,\GeV$. We show the result obtained
  with a pure hadron gas equation of state (HG-EoS) and the combination
  of the hadron gas together with a lattice EoS at higher temperatures
  (HG+LAT-EoS). In both cases the hadronic rates are used up to
  $T=170\,\MeV$ and partonic rates for higher termperatures. The model
  results are compared to the experimental data obtained by the STAR
  Collaboration \cite{Adamczyk:2015lme}.}
\label{dilRHICcompEoS}
\end{figure}
\begin{figure}
\includegraphics[width=1.0\columnwidth]{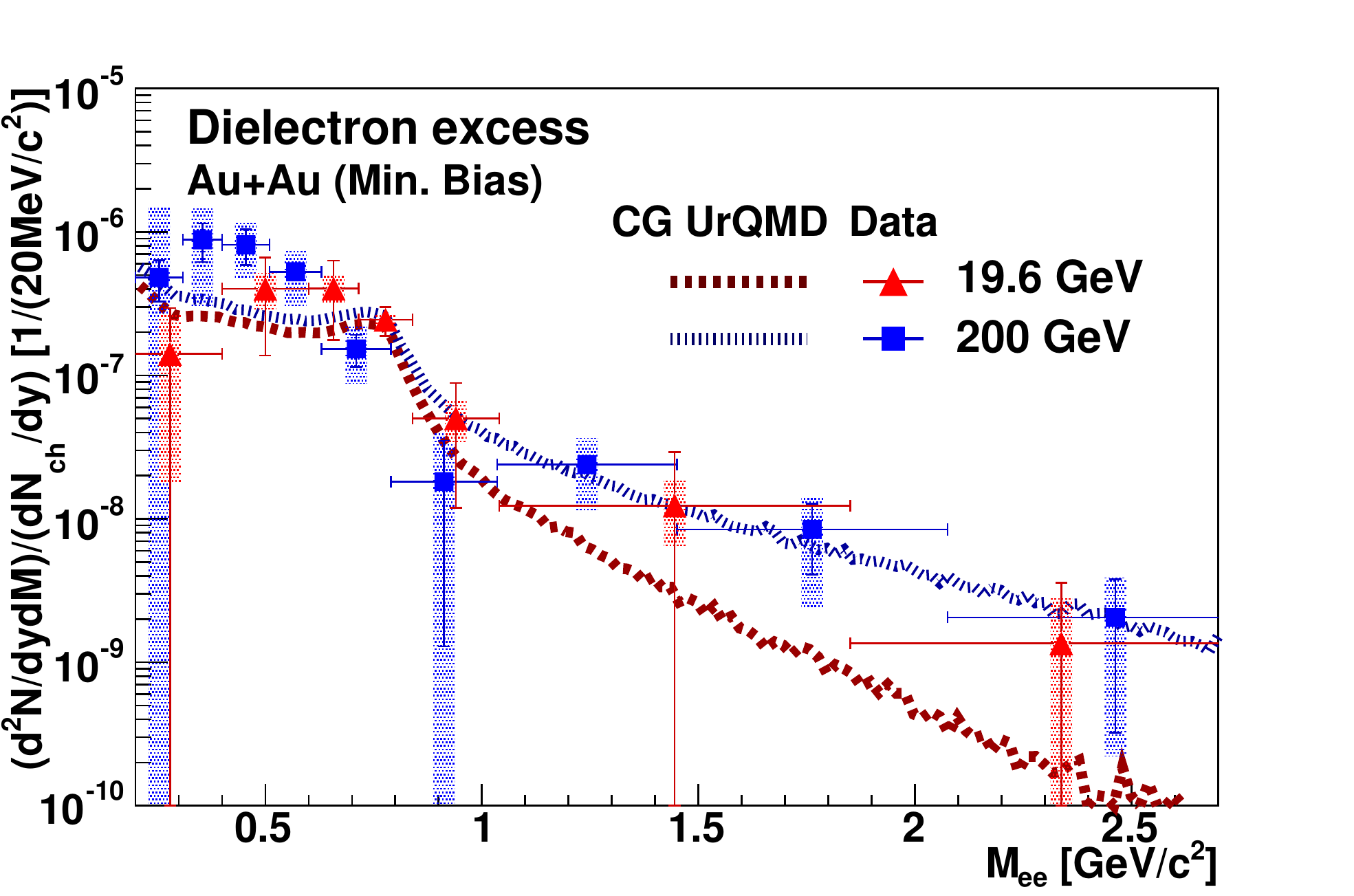}
\caption{(Color online) Dielectron excess spectrum for minimum bias
  (0-80\% most central) Au+Au collision at $\sqrt{s_{NN}}=19.6$ and
  $200$\,GeV as obtained with the coarse-graining of UrQMD simulations
  (CG UrQMD). The results include the thermal contributions from the
  $\rho$, multi-pion interactions and the QGP. Additionally the UrQMD
  freeze-out $\rho$ is included for this calculation. The model results
  are compared to the results of experimental measurements by the STAR
  Collaboration \cite{Adamczyk:2015mmx}.}
\label{dilexcess}
\end{figure}

It is interesting that the spectral shape of the
thermal $\rho$ resembles its vacuum shape in both cases, compared to the
very strong broadening and low-mass enhancement which is observed for
SIS\,18 and FAIR energies
\cite{Endres:2015fna, Endres:2015egk}. 
However, this is not surprising
since in the previous section it has already become clear that the
baryochemical potential is rather low in most of the cells. And even if
one considers that the baryonic modifications of the spectral shape for
the $\rho$ are governed by the effective baryon and anti-baryon density,
the effects seem relatively small. One reason for this is that the
initial heating is faster and stronger at RHIC energies and the early
phase of the reaction is mostly dominated by partonic emission (which is
quite insensitive with regard to finite quark chemical potential
$\mu_{q}=1/3\mu_{\mathrm{B}}$), whereas the hadronic contributions are
predominantly radiated at later stages when the baryon densities are
lower. Consequently, the baryon-induced medium effects---which are the
main cause of the $\rho$ low-mass enhancement---are only very moderate here. Note
that there is also a significant non-thermal $\rho$ contribution from
low-temperature and late-stage cells, which is more dominant for
200\,GeV. This might be due to the longer lifetime of the system, with a
significant number of those mesons in peripheral cells and late in
the evolution. In contrast to the thermal $\rho$, the thermal $\omega$
contribution is rather negligible compared to the respective freeze-out
contribution. This is mainly due to the long lifetime of the $\omega$,
which is typically so long that this resonance mostly decays outside the hot and
dense region.

In contrast to the low-mass region, for
$M_{\mathrm{e}^{+}\mathrm{e}^{-}}> 1\,\GeV/c^{2}$ the overall dilepton
yield is no longer dominated by the peaks from various hadronic decays but one
experimentally finds a structureless continuum. In our model the thermal
emission from multi-pion interactions and from the partonic phase shine
in this part of the spectrum. Note however that---as mentioned before---
the present calculation does not include the Drell-Yan and, more
important, the open-charm contributions to the spectrum. Nevertheless,
as the strength of possible medium modification for $D$ or $\bar{D}$
mesons is yet unclear, our calculation can serve as a thermal baseline.

For $\sqrt{s_{NN}}=19.6$\,GeV the QGP emission is the dominant
contribution in the mass region from 1 to $2.8\,\GeV/c^{2}$ with a
significant contribution from the multi-pion part which is strongest
around $M_{\mathrm{e}^{+}\mathrm{e}^{-}} = 1.1$\,GeV$c^{2}$. Here
20-30\% of the thermal contribution are from the hadronic source, while
for higher masses the multi-pion yield becomes rather insignificant. The
comparison with experimental data allows no clear conclusions at this
energy due to the limited statistics and rather large errors. The yield
from the coarse-graining model is within the statistical error of the
data but rather at the lower boundary. The situation is somewhat
different for Au+Au collisions at 200\,GeV. At this higher energy the
QGP emission is now the dominant thermal contribution, whereas the
hadronic contribution is suppressed by at least a factor of 10. Due to
the significantly better statistics, one can observe that the model does
not fully describe the STAR data, but the dilepton emission obtained
within the model makes up for only roughly 50\% of the measured yield in
the region from 1 to 2\,GeV/$c^{2}$. Interestingly, for even higher
masses the agreement between model an data becomes better, the slope of
the thermal emission seems to be slightly harder than the measured
one. These results agree with previous studies indicating that the
relative suppression of the charm contribution due to medium effects is
more pronounced at higher masses, leaving it the dominant contribution
only for lower masses around 1\,GeV/$c^{2}$ \cite{Lang:2013wya}.
\begin{figure*}
\includegraphics[width=1.0\columnwidth]{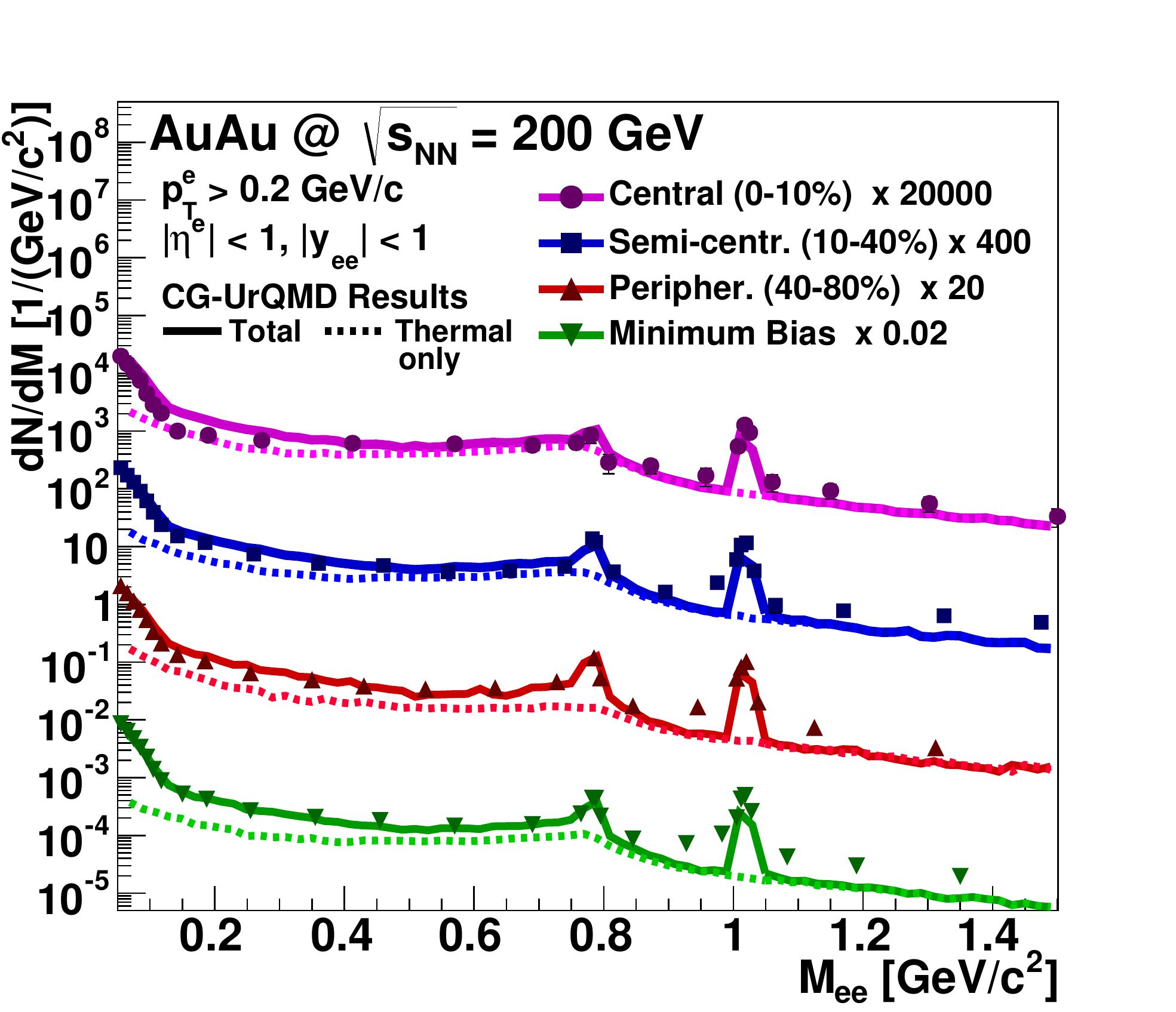}
\includegraphics[width=1.0\columnwidth]{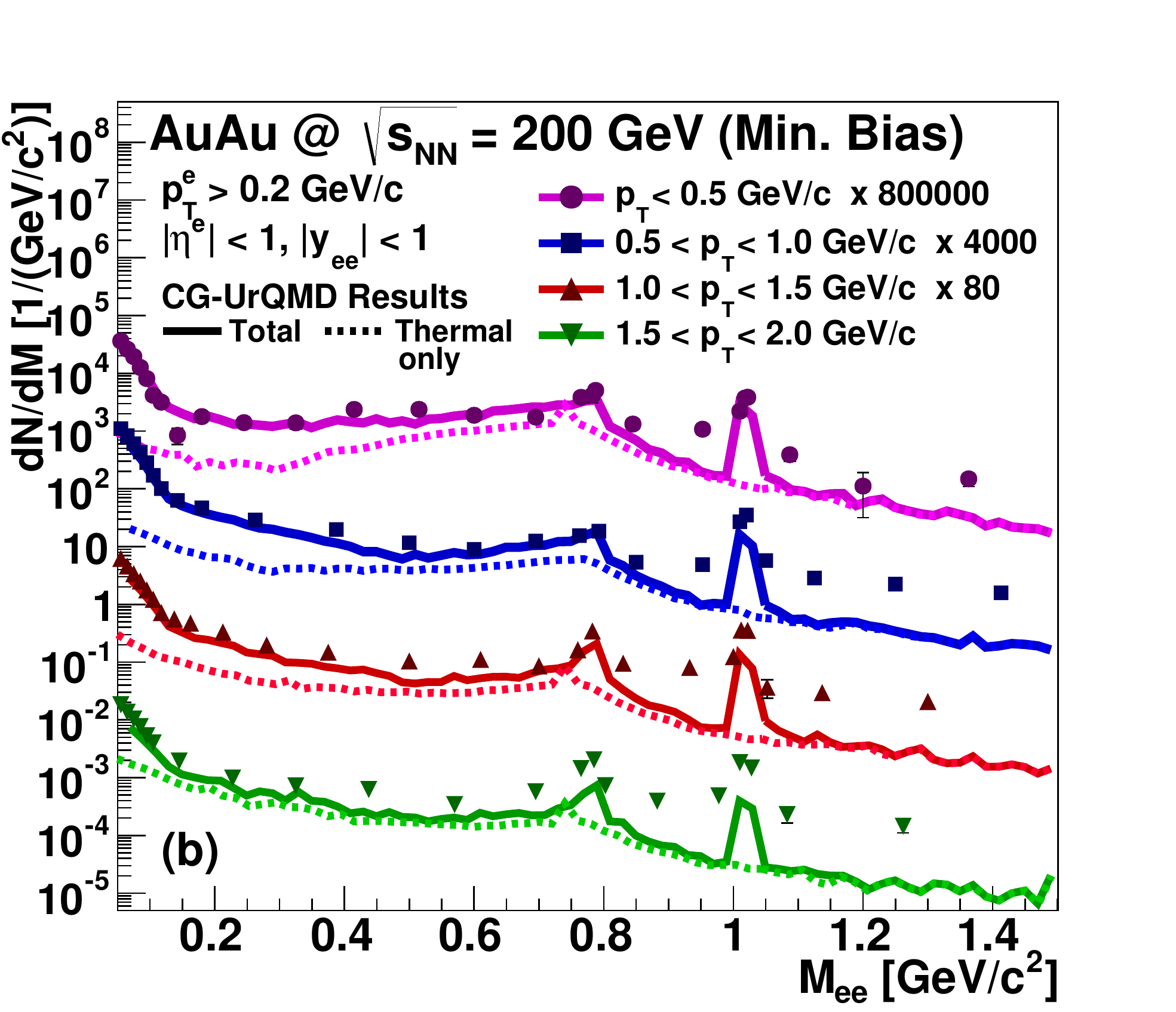}
\caption{(Color online) Dielectron invariant-mass spectra for Au+Au
  collisions at $\sqrt{s_{NN}}=200$\,GeV for different centralities (a)
  and transverse-momentum bins (b). Besides the total yields (full
  lines) we also show the thermal contribution separately (dashed
  lines). The results in (a) are shown for central (0-10\%),
  semi-central (10-40\%), peripheral (40-80\%), and minimum bias (i.e.,
  0-80\%) events. In (b) the spectra are shown for minimum bias
  collisions in four transverse-momentum bins covering the range from
  $p_{t}=0$ to 2.0\,GeV/$c$. The spectra are shown within the STAR
  acceptance and compared to the experimental data
  \cite{Adamczyk:2015lme}. In addition, they are scaled for better
  comparability.}
\label{RHICcentrpt}
\end{figure*}

While by default we use a combination of a hadron gas and a lattice EoS
(HG+Lat-EoS) for all calculations presented in this work, it was
discussed in Sec.\,\ref{ssec:CGapproach} that this is not fully
consistent with the underlying purely hadronic microscopic dynamics. In
consequence, it is instructive to compare this standard scenario with
the more consistent case where only the hadron gas equation of state
(HG-EoS) is used for all temperature ranges to extract $T$ and
$\mu_{\mathrm{B}}$. Note that in both cases we use the hadronic rates up
to $T=170$\,MeV and the partonic emission rates for higher temperatures
for being able to directly compare the effect of the different EoS. The total
invariant-mass spectra obtained with both EoS are put on top of each other for
minimum bias Au+Au collisions at 200\,GeV in
Fig.\,\ref{dilRHICcompEoS}. The results
indicate that the differences with regard to the overall yield in the low mass region are rather small and
result in no significant deviations in the thermal emission pattern for
masses up to $M_{\mathrm{e^{+}e^{-}}}=1\,\GeV/c^{2}$. The slightly
reduced QGP yield in this region due to the lower temperatures from the
HG-EoS is mostly compensated by a larger hadronic contribution, especially around the $\rho$ pole mass. However, the
picture is quite different for masses above $1\,\GeV/c^{2}$, dominated
by the QGP emission: Here the use of the HG-EoS results in a significantly lower
thermal yield and a softer slope. The yield is suppressed by almost an
order of magnitude at $M_{\mathrm{e^{+}e^{-}}}=2.5\,\GeV/c^{2}$ compared
to the HG+Lat-EoS scenario. This is not surprising, as these higher
masses are dominated by emission from the very early hot stage of the
fireball where the highest energy densities are reached. Here the
differences between the two EoS are most dominant and the lattice
equation of state results in significantly higher temperatures. On the
one hand, this result indicates that the low-mass dilepton spectra are
quite insensitive with regard to the EoS; on the other hand, it shows
again that direct information regarding the phase structure of QCD might
be deduced from the spectra at higher invariant masses,
$1\;\GeV/c^2 \lesssim M_{\mathrm{e^{+}e^{-}}} \lesssim 2.5 \;\GeV/c^2$. However, the
experimental extraction of the thermal yield is difficult in this region
as also a strong contribution from correlated charm decays is found
here, see discussion above.

In addition to the full invariant-mass distributions, the STAR
Collaboration also published dilepton excess spectra for minimum bias
Au+Au collisions at 19.6 and 200\,GeV \cite{Adamczyk:2015mmx}. Here the
cocktail contributions (hadronic decays, Drell-Yan and open charm) are
subtracted such that the resulting spectra respresent only the thermal
dilepton emission. Furthermore the data are corrected for the
experimental acceptance. In Figure~\ref{dilexcess} these results are
compared to the thermal contribution from our model, including the
non-thermal UrQMD ``freeze-out'' $\rho$ and excluding the thermal
$\omega$ contribution. (The $\omega$ is usually treated as part of the
cocktail and was subtracted from the experimental spectrum.) We see that
for the mass region $M_{\text{e}^{+}\text{e}^{-}}>1\,\GeV/c^{2}$ in Au+Au collisions at
$\sqrt{s_{NN}}=200$\,GeV the thermal result agrees very well with the
data, indicating that the thermal part of this mass region seems to be
accurately described with the coarse-graining approach. However, note
that the subtracted cocktail contribution does not account for medium
modifications of the charm contribution, so that the meaning of the
high-mass excess spectrum is rather limited. At 19.6\,GeV the thermal
spectrum from the model seems to be slightly below the data for higher masses, but still
within the large statistical and systematic errors. In the low-mass
region the agreement between model and data is better for 19.6\,GeV than for 200\,GeV, but in both cases the experimental thermal excess seems to be
slightly underestimated by the model. Nevertheless, considering the
uncertainty of the data and the subtraction procedure the agreement is
quite satisfactory.

\begin{figure}[b]
\includegraphics[width=1.0\columnwidth]{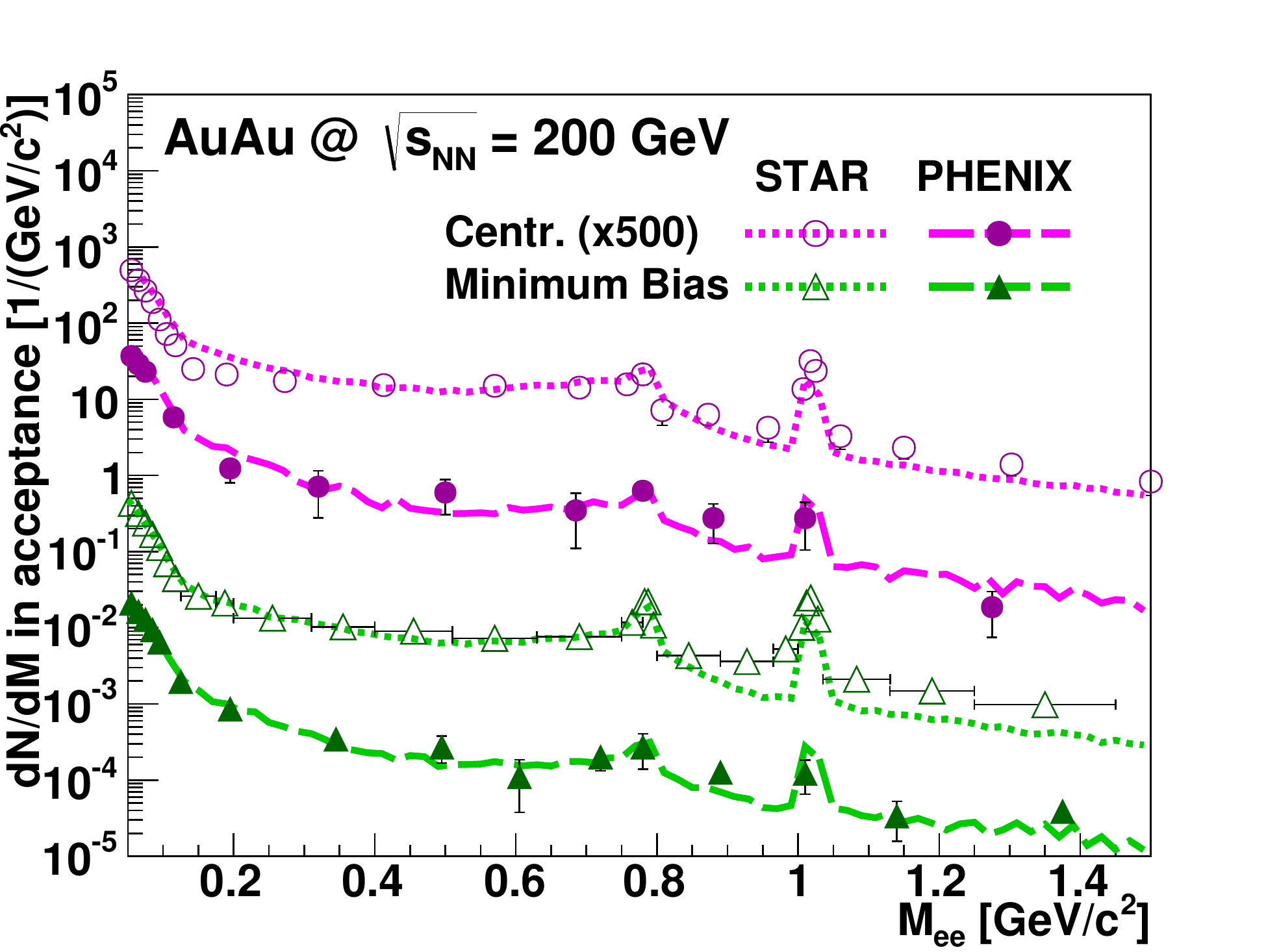}
\caption{(Color online) Comparison of the dielectron invariant-mass
  spectra for central (magenta) and minimum bias (green) Au+Au
  collisions at $\sqrt{s_{NN}}=200$\,GeV with the STAR and PHENIX data
  \cite{Adare:2015ila, Adamczyk:2015lme} in the corresponding
  acceptances. The total sum of the model results for PHENIX is given by
  the long dashed lines, and for STAR by the short dashed lines.}
\label{STARPHENIX}
\end{figure}

So far we have considered dilepton spectra for minimum bias reactions
and the full transverse-momentum range, but the thermal dilepton yield also largely depends on
the centrality of the reaction and on the transverse-momentum window in
which the results are measured. Both dependencies were investigated by
the STAR Collaboration for Au+Au collisions at 200\,GeV
\cite{Adamczyk:2015lme}, and the experimental data are presented
together with the model results in Figure~\ref{RHICcentrpt}. The left
plot (a) shows the invariant-mass spectra for central (0-10\%),
semi-central (10-40\%) and peripheral (40-80\%) collisions, together
with the minimum bias result from Fig.~\ref{dilinvmassRHIC}\,(b) for
comparison. In all four centrality classes one observes quite a good
description of the low invariant-mass data by the coarse-graining
results. For higher masses larger than $1\,\GeV/c^{2}$ the
underestimation of the dilepton yield observed for minimum bias reactions is also
found for other centrality classes. However, for the most central
reactions the description seems to be slightly better. In this case the thermal
emission alone can almost fully describe the dilepton data for higher masses. This would
be in accordance with the assumption that the medium effects on the open
charm production are most dominant for central collisions, leading to a
suppression of the open-charm contribution to the dilepton spectra.

For the $p_{t}$ dependence of $\mathrm{e}^{+}\mathrm{e}^{-}$
production, the comparison between theory and data gives a more nuanced
picture, as presented in Figure~\ref{RHICcentrpt}\,(b). Here the
scaled results for minimum bias Au+Au collisions in four different
transverse momentum bins are shown. In the low invariant-mass region one
finds a good description of the data for the lower transverse momentum
bins up to $1\,\GeV/c$, while especially for $p_{t} > 1.5\,\GeV/c$ the
measured results are underestimated by up to a factor 2. Interestingly,
this does not only affect the thermal yield, but also the pure
hadronic cocktail contributions, as can be seen from the underestimate
for the $\pi$-dominated very low masses and the $ \omega$ and $\phi$
peaks. The reason for this might be the expansion dynamics from
the underlying transport model, which is known to somewhat underestimate
the collective flow of the fireball \cite{Petersen:2006vm}, resulting in too soft $p_{t}$
spectra for the produced particles. However, the general trend when
going from the low to the high transverse momentum region is the
increasing importance of the thermal emission in the low-mass region and
a flattening of the shape of the spectrum. This is due to two effects:
On the one hand, the $p^{\mathrm{e}}_{t}$ cut for single electrons leads to a
suppression of low masses ($M < 0.4 \,\GeV/c^{2}$) when the transverse
momentum of the pair is close to zero. On the other hand, the emission
of high-$p_{t}$ dileptons occurs mostly at the higher temperatures which
can be found in the early Quark-Gluon Plasma phase, whereas the hadronic
emission is usually found to be softer.

Regarding the higher invariant-mass region for $M > 1 \,\GeV/c^{2}$, an
underestimation of the thermal yield is visible, reaching from a factor
2 for low $p_{t}$ up to a factor of 10 for the higher transverse
momenta. This underprediction is not surprising, as it was already visible
in the full $p_{t}$-integrated invariant-mass spectrum. As mentioned
above, this is clearly due to the absence of the charm and Drell-Yan
contributions in our calculation.

Although we have up to this point focused on the measurements by the
STAR Collaboration, it is natural to compare the model results obtained
from the coarse-graining approach also with the results of the PHENIX
Collaboration. This is of importance, as the first results from PHENIX
showed a strong enhancement of the dilepton invariant-mass spectrum for
$0.3 < M_{\text{ee}} < 0.7\,\GeV/c^{2}$ in central collisions, which was
not compatible with the results from the STAR Collaboration
\cite{Adare:2009qk}. In consequence, there has been much discussion
about the different detector properties and corresponding acceptances,
which made a direct comparison of the two results difficult. Also
theoretical models failed to reproduce the PHENIX results
\cite{Rapp:2010sj, Linnyk:2011vx}. Recently, the PHENIX Collaboration
published new results measured with an updated experimental set-up,
including a hadron-blind detector (HBD) which could significantly
improve the electron identification and the signal sensitivity
\cite{Adare:2015ila}. In Figure~\ref{STARPHENIX} we show the model
results for both central and minimum bias Au+Au collisions at
$\sqrt{s_{NN}}=200$\,GeV within the PHENIX and STAR acceptances,
together with the corresponding experimental data. The comparison
clearly shows that the model not only describes the STAR data, but also
the new PHENIX results for central as well as minimum bias
collisions. However, note that the statistics obtained by PHENIX is
significantly lower, leading to larger errors of the measurement. The
main explanation for this is the two-arm set-up of the PHENIX detector
so that many of produced electrons and positrons do not reach the
detector; if only one particle of a pair reaches the detector, this
further increases the background of the measurement. Nevertheless,
within the errors of the measurement one can state that the PHENIX and
STAR dilepton measurements now fully agree with each other and that the
low-mass excess above the hadronic cocktail can be explained by thermal
hadronic and partonic emission from medium-modified spectral functions.

To conclude the study for RHIC energies, the model results are finally
compared to the transverse momentum spectra from the PHENIX measurement
in Figure~\ref{PHENIXpt}. The (scaled) data and model results within experimental
acceptance are presented for three different invariant-mass bins. The
thermal contribution and the hadronic decay cocktail from UrQMD are
shown separately, as well as the total yield. At very low masses
($M<0.1\,\GeV/c^{2}$) the hadronic cocktail contribution dominates the
dilepton emission, mainly stemming from $\pi^{0}$ decays. Only for high
$p_{t}$ larger than $1.5\,\GeV/c$ the thermal emission becomes
significant. However, such high momenta are largely suppressed by a
factor of $100$ in that mass region. The model results agree quite well
with the experimental measurements, only for lower $p_{t}$ a slight
overestimation of the yield is obtained. (Note that dilepton pairs with
$p^{\text{ee}}_{t}<0.4\,\GeV/c$ are out of the PHENIX acceptance in this
mass bin, as the single electron transverse momentum is required to be
larger than $0.2\,\GeV/c$.) In the mass region from 0.3 to
$0.76\,\GeV/c^{2}$ the thermal and nonthermal emission almost equally contribute for
low $p_{t}$ with a slight dominance of the hadronic cocktail for
transverse momenta from 0.5 to $1.0\,\GeV/c$. In contrast, the thermal
dilepton emission clearly outshines the hadronic decays for higher
$p_{t}$ values above 1.5\,GeV/$c$. Note that the present findings from the
coarse-graining approach for this mass region roughly agree with the results
from a fireball parametrization (using the same spectral functions as in
our model) where the non-thermal emission dominates for lower momenta and the
thermal contribution---mainly from the $\rho$---for higher momenta
\cite{Adare:2015ila}. For the mass region
$M_{\mathrm{e}^{+}\mathrm{e}^{-}}>1.2\,\GeV/c^{2}$ the thermal emission
(i.e, here almost exclusively the partonic contribution from the QGP) is
clearly the dominant source in the present calculations. However, the yield obtained with the coarse-graining approach is
below the data about a factor 2-3 for low $p_{t}$ and up to 10 for
higher momenta, once again indicating the missing contributions from
open-charm mesons. In spite of the significantly differing acceptances of the STAR and PHENIX experiments, the present results are consistent with the findings from the comparison of model results and data for the invariant-mass spectra in various $p_{t}$-bins (see Fig.~\ref{RHICcentrpt}\,(b)).

\subsection{\label{ssec:LHC} Large Hadron Collider (LHC)} 
\begin{figure}
\includegraphics[width=1.0\columnwidth]{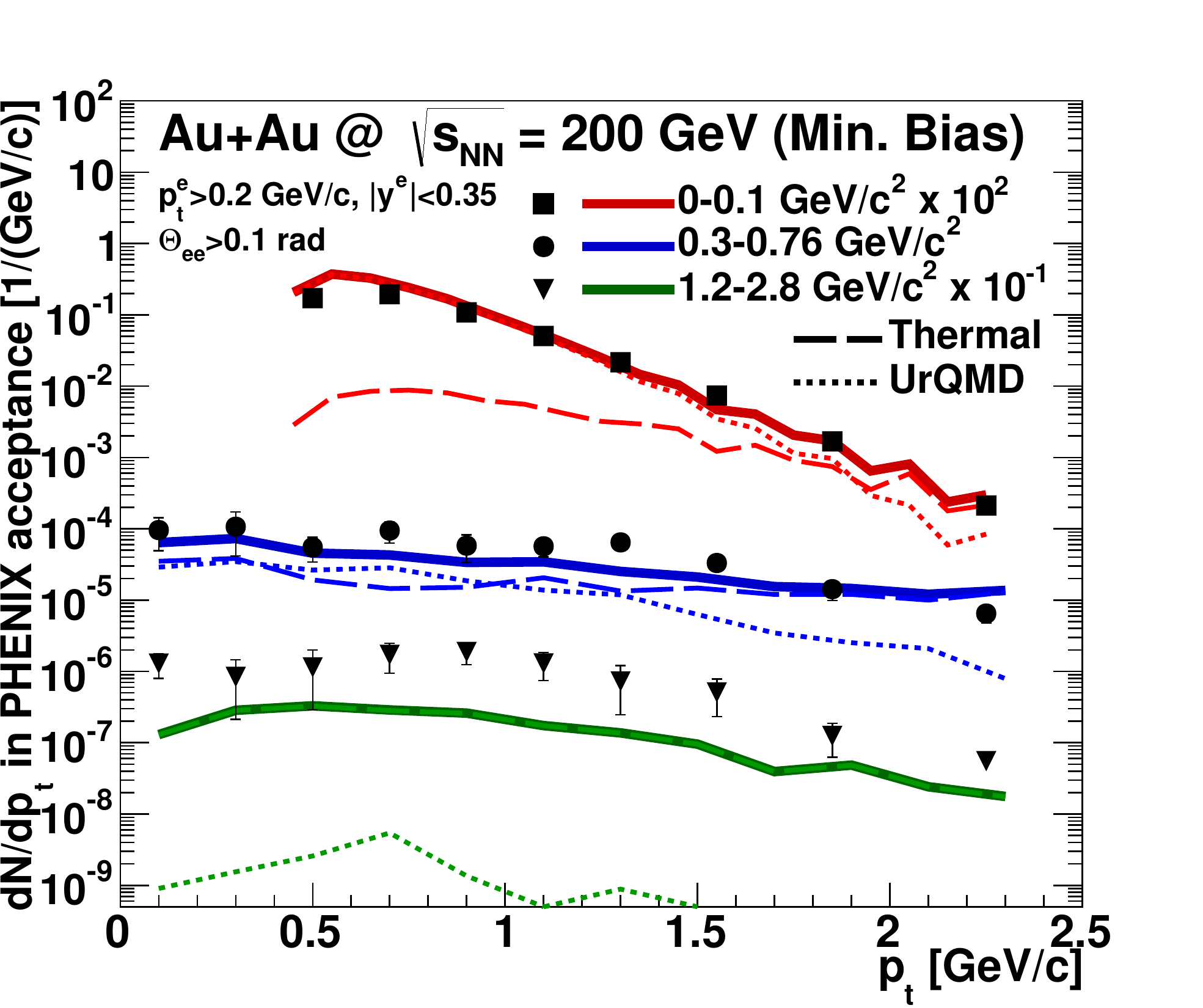}
\caption{(Color online) Dielectron transverse-momentum spectra for three
  mass bins (red: $M=0-0.1\,\GeV/c^{2}$, blue: $M=0.3-0.76\,\GeV/c^{2}$,
  blue: $M=1.2-2.8\,\GeV/c^{2}$) within the PHENIX acceptance. The
  results here are for minimum bias Au+Au collisions 
  at$\sqrt{s_{NN}}=200$\,GeV. Besides
  the total yields from the model calculations (full lines) also the
  thermal (long dashed) and non-thermal hadronic decay contributions
  (short dashed) are presented. For comparison the experimental data
  from the PHENIX Collaboration \cite{Adare:2015ila} are shown as well.}
\label{PHENIXpt}
\end{figure}
\begin{figure*}
\includegraphics[width=1.0\columnwidth]{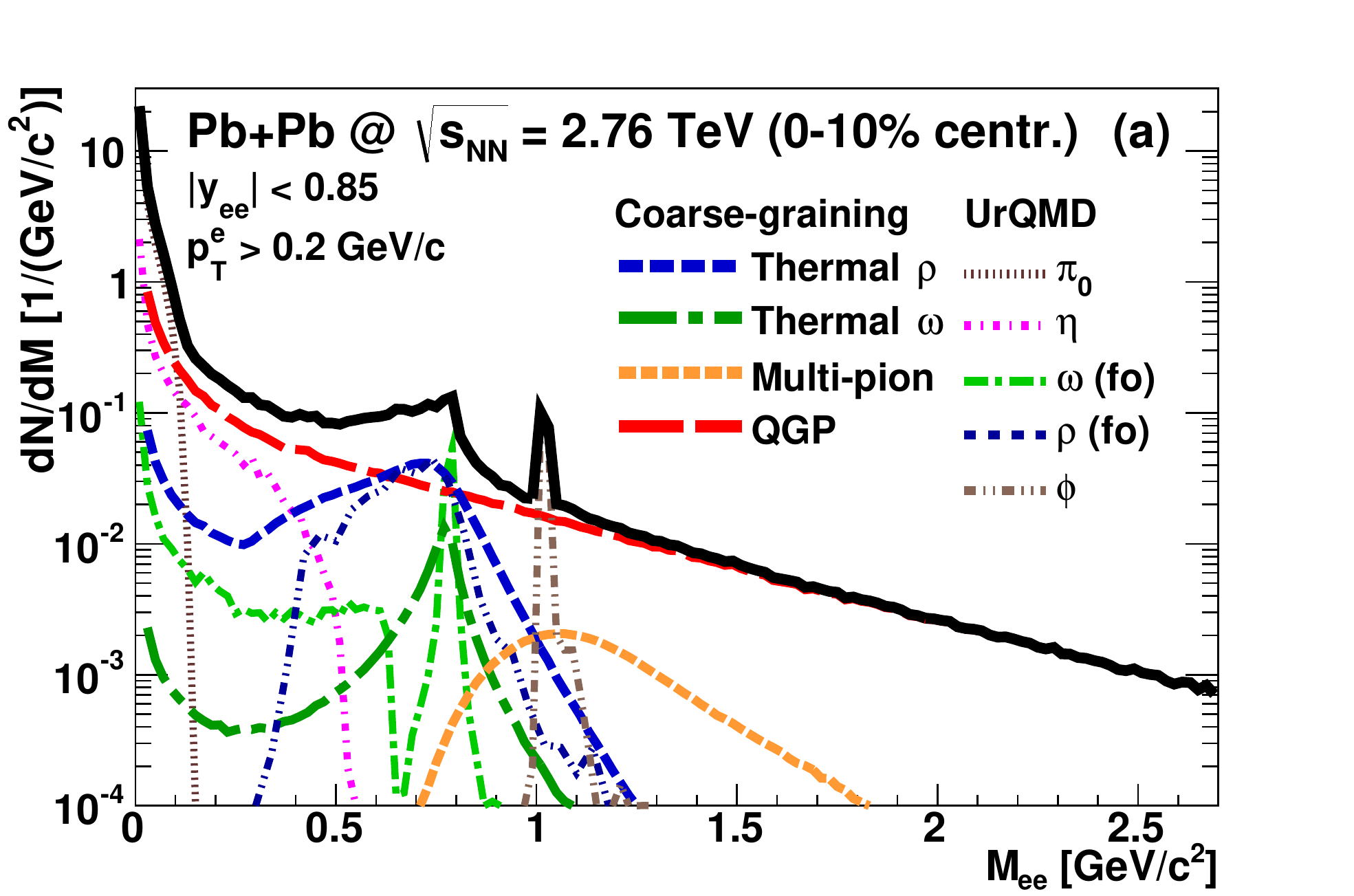}
\includegraphics[width=1.0\columnwidth]{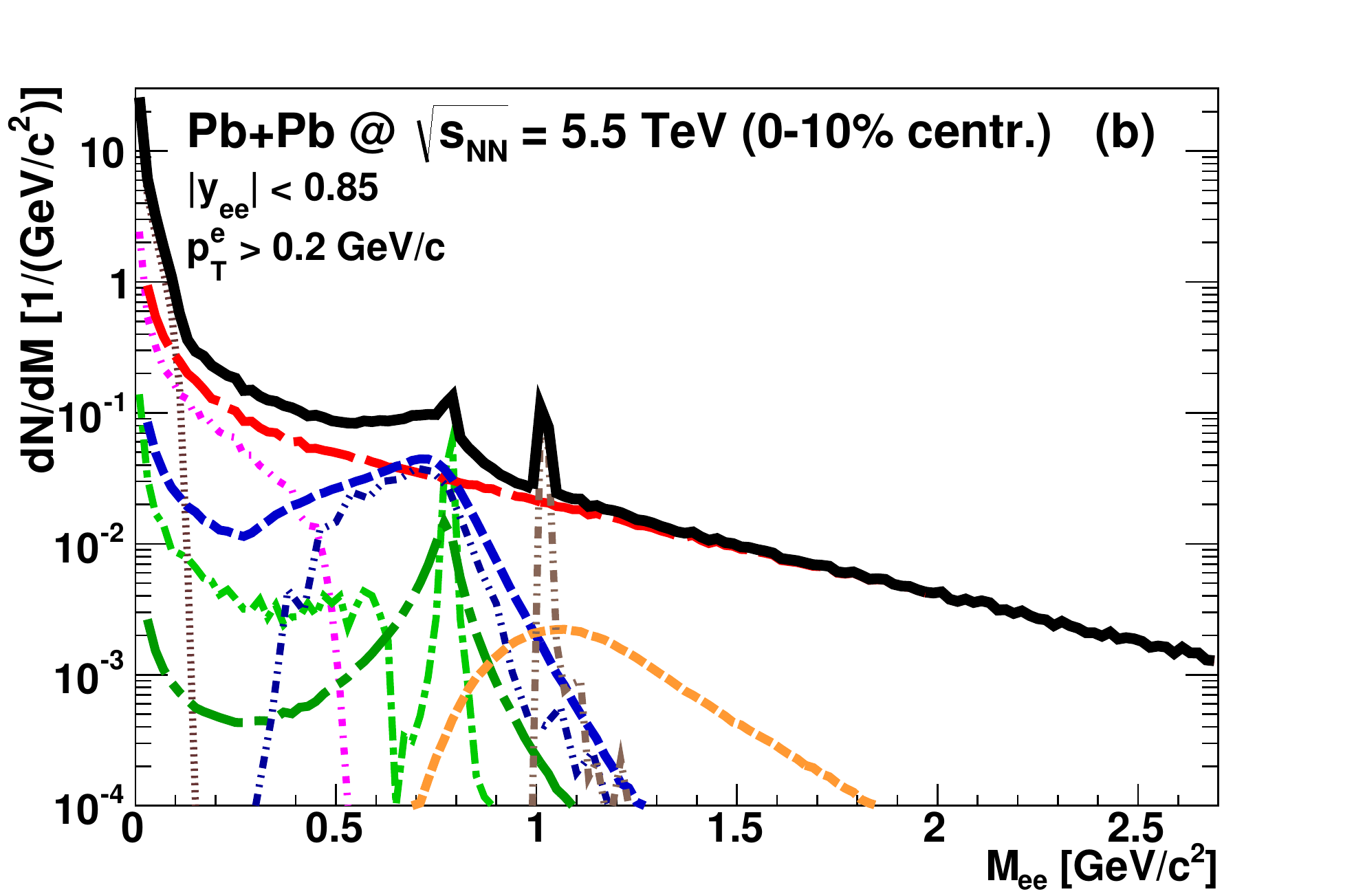}
\caption{(Color online) Dielectron invariant-mass spectra for 0-10\% most
  central Pb+Pb collisions at $\sqrt{s_{NN}}=2.76\,\TeV$ (a) and
  $5.5\,\TeV$ (b). The sum includes the thermal hadronic and partonic
  emission obtained with the coarse-graining approach, and also the
  hadronic $\pi$, $\eta$, and $\phi$ decay contributions from UrQMD as
  well as the ``freeze-out'' contributions (from cold cells) of the
  $\rho$ and $\omega$ mesons.}
\label{dilinvmassLHC}
\end{figure*} 
\begin{figure}[b]
\includegraphics[width=1.0\columnwidth]{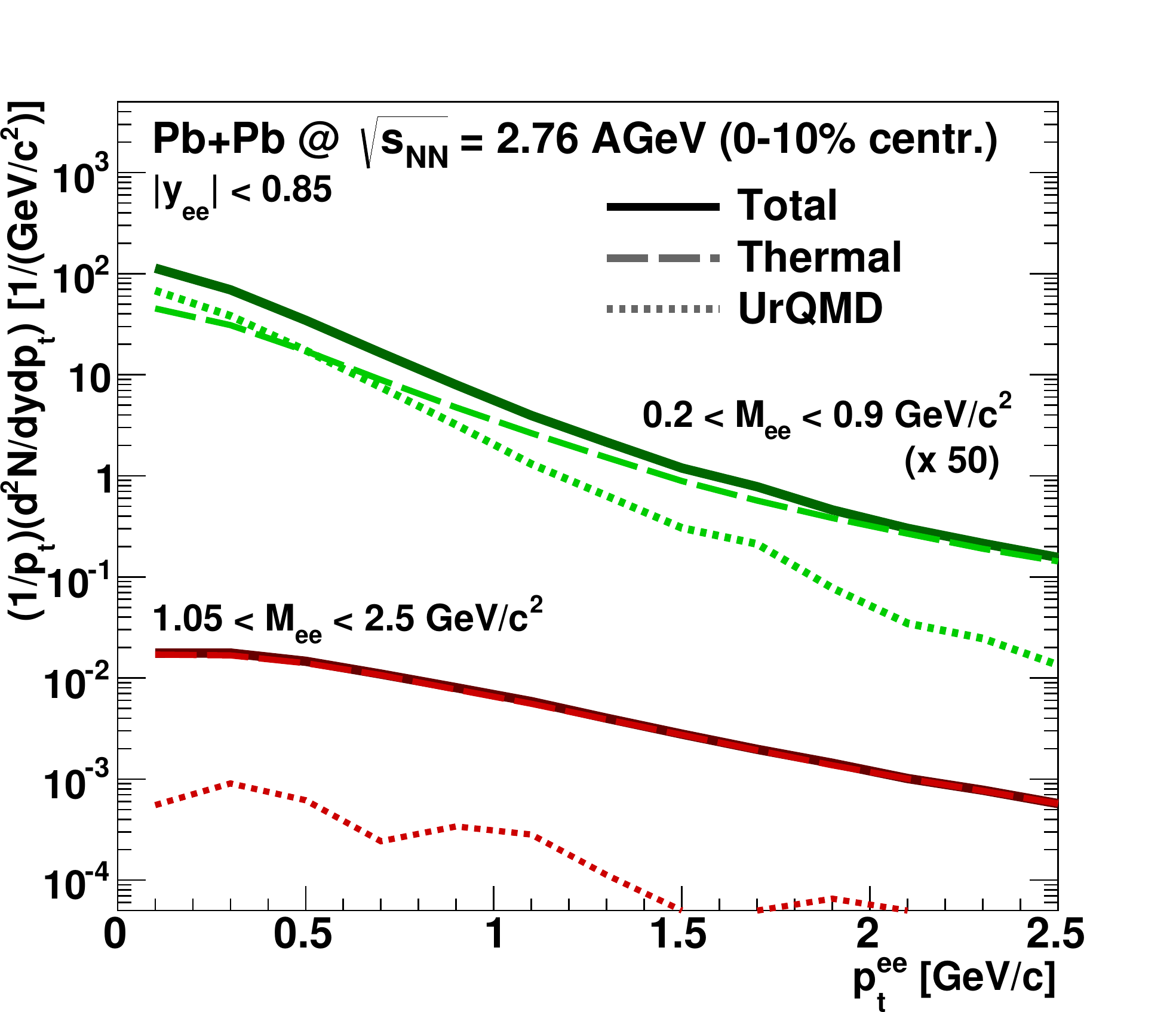}
\caption{(Color online) Dielectron transverse-momentum spectra for
  central Pb+Pb collisions at $\sqrt{s_{NN}}=2.76$\,TeV. The results are
  shown for the low-mass
  ($0.2< M_{\mathrm{e}^{+}\mathrm{e}^{-}}<0.9\,\GeV/c^{2}$; green) and
  intermediate mass region
  ($1.05< M_{\mathrm{e}^{+}\mathrm{e}^{-}}<2.5$\,GeV/$c^{2}$;
  red). Besides the total yields from the model calculations (full
  lines) also the thermal (long dashed) and non-thermal UrQMD hadronic
  decay contributions (short dashed) are presented.}
\label{LHCpt}
\end{figure} 
\begin{figure*}
\includegraphics[width=1.0\columnwidth]{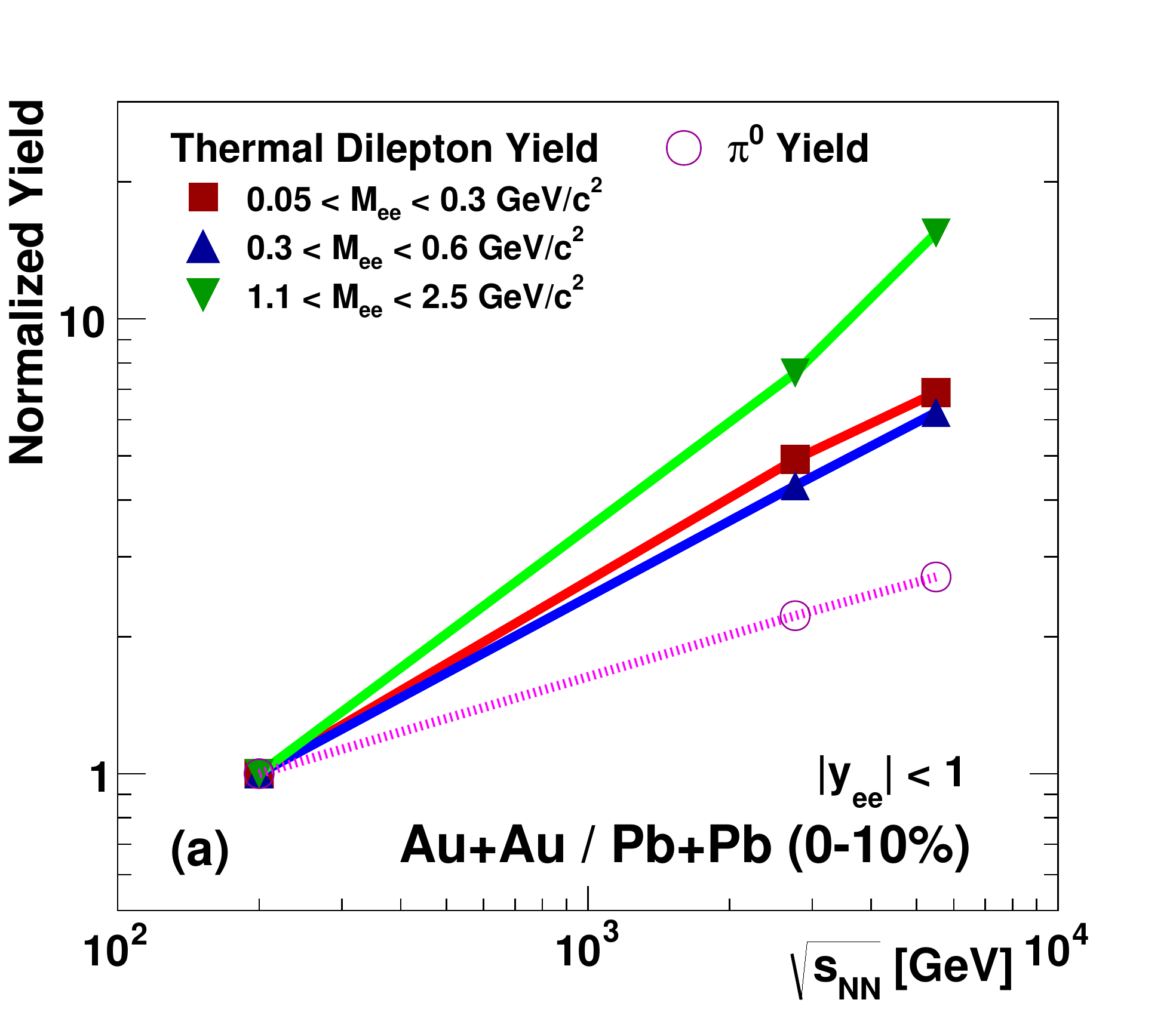}
\includegraphics[width=1.0\columnwidth]{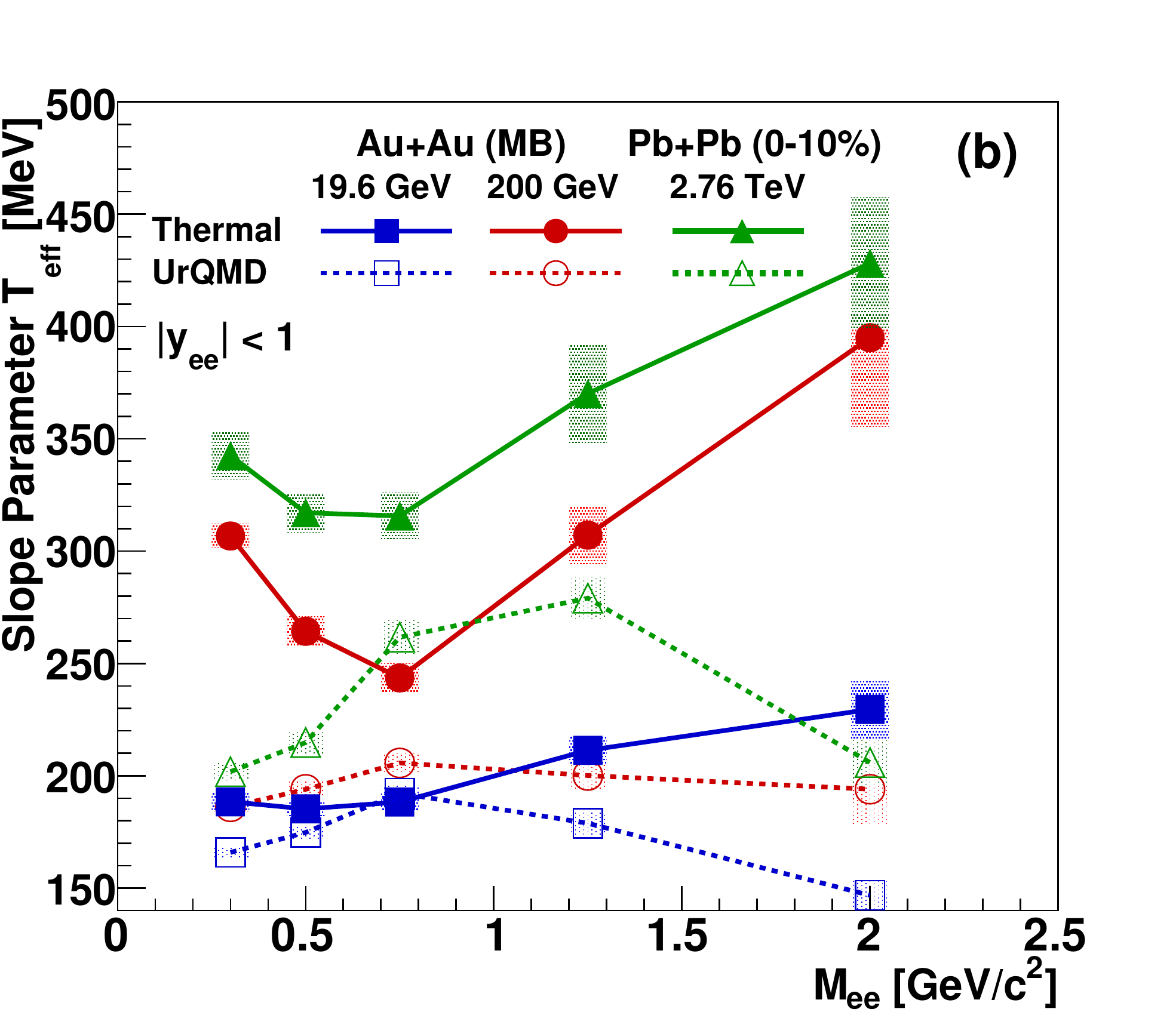}
\caption{(Color online) (a) Thermal dilepton yield for three different
  mass regions and the $\pi^{0}$ yield for central Au+Au/Pb+Pb
  collisions, normalized to the result for $\sqrt{s_{NN}}=200$\,GeV. (b)
  Mass dependent slope parameter $T_{\mathrm{eff}}$ for the thermal
  (full lines) and non-thermal (dashed lines) dilepton yields in minimum
  bias Au+Au collisions at 19.6\,GeV (blue) and 200\,GeV as well as for
  central Pb+Pb at 2.76\,TeV (green). The error bands indicate the
  systematic error of the fit. All results are for mid-rapidity, i.e., for $|y_{\text{ee}}| < 1$.}
\label{slope}
\end{figure*}
As became clear from the study of the reaction dynamics in
Sec.~\ref{ssec:fireball}, the fireball of hot and dense matter created
in a heavy-ion collision reaches higher temperatures and cools down
more slowly at the Large Hadron Collider in comparison to the reaction
evolution for RHIC energies. However, there is no significant change
with regard to the baryon densities, the baryochemical potential was
already close to zero for most cells at
$\sqrt{s_{NN}}=200$\,GeV. Consequently, the resulting invariant-mass
spectra for central (0-10\%) Pb+Pb collisions at center-of-mass energies
of 2.76 and 5.5\,TeV as shown in Figure~\ref{dilinvmassLHC} exhibit the
same mostly vacuum-like spectral shape of the $\rho$ meson contribution,
together with an increased yield stemming from the Quark-Gluon
Plasma. The partonic contribution is dominating the spectra for masses
above $0.1\,\GeV/c^{2}$, except for the pole-mass peaks of the three
vector mesons $\rho$, $\omega$ and $\phi$. However, the $\rho$
contribution still plays a significant role as well, and its relative
strength is not much smaller than at the top RHIC energy. In general, the
increasing number of hot cells with temperature above $T_{c}$ goes along
with a longer lifetime of the fireball and a larger overall thermal
four-volume also for temperatures below the critical temperature, which
equally leads to a rise of the hadronic emission. In the same manner as
there is no strong change of the spectra from RHIC to LHC, the situation
does hardly change when going from 2.76\,TeV to the maximum LHC energy
of 5.5\,TeV, except for an higher overall yield. We will study this
energy dependence in more detail in Sec.\,\ref{ssec:Comp}.

The transverse momentum spectra for 2.76\,TeV in Figure~\ref{LHCpt} are
shown in two different mass bins, for the low-mass region
($0.2 < M_{\text{ee}} < 0.9\,\GeV/c^{2}$) and for the intermediate
masses above the $\phi$ pole mass
($1.05 < M_{\text{ee}} < 2.5\,\GeV/c^{2}$). For the lower masses the
finding is similar to those for Au+Au collisions at 200\,GeV (compare
Fig.\,\ref{PHENIXpt}): The hadronic sources are more dominant at low
$p_{t}$, while the thermal emission is the major contribution for high
momenta. In general the slope of the thermal emission is harder (i.e.,
flatter) than that of the hadronic decays. For the intermediate mass
region above $1\,\GeV/c^{2}$, the only dominant contribution stems from
the Quark-Gluon Plasma, whereas the hadronic decays become
negligible. The overall slope of the higher masses is also harder,
indicating emission from hotter cells on the one hand, but also the
stronger flow which is proportional to the mass of the particles. As
before one should, however, bear in mind that a full study for the high
masses would need to include the missing charm and Drell-Yan
contributions.

\subsection{\label{ssec:Comp} Comparison of RHIC and LHC results} 

The previous results have already shown that the temperature and
lifetime of the fireball increase when going from RHIC to LHC energies,
which is connected with a larger yield from thermal dilepton
production. In the following, these very qualitative findings shall be
investigated in more detail.

In Figure~\ref{slope}\,(a) the relative ratio of the thermal dilepton
yield at mid-rapidity ($|y_{\text{ee}}|<1$) for different mass regions is shown in relation to the yield which
is obtained for Au+Au collisions at $\sqrt{s_{NN}}=200$\,GeV. In
addition, the increase of the $\pi^{0}$ yield is shown, as comparison of
the thermal results with the production of hadrons or dileptons from
hadronic decays, respectively. The results depict that in general the
thermal contributions exhibit a stronger increase than the $\pi^{0}$
yield. For the lower masses---0.05 to 0.3 and 0.3 to
$0.6\,\GeV/c^{2}$---the thermal yield scales with the number of neutral
pions as $N_{\pi^{0}}^{\alpha}$, with $\alpha$ found to be approximately
1.9 here. For the mass region above the $\phi$ pole mass, where purely
the QGP contributes to the thermal emission, the relative increase is
even stronger with $\alpha \approx 2.4$. Note that the exponent $\alpha$
for the mass region where the excess above the cocktail is found (i.e.,
0.3-$0.6\,\GeV/c^{2}$) is similar and only slightly larger compared to
the one obtained using a fireball parametrization \cite{Rapp:2013nxa};
there the scaling with the total number of charged hadronic particles is
found to be $N_{\mathrm{ch}}^{\alpha}$ and $\alpha=1.8$. The somewhat
stronger enhancement of the high-mass yield at LHC energies is explained by the fact that the number of QGP-emitting hot cells exhibits a larger
increase than the lower temperature four-volume (compare Figs.\,1 and
2).

Whereas the thermal yields alone allow for only rather qualitative
conclusions regarding the underlying reaction dynamics, another
observable that helps to determine the temperature and expansion
dynamics of the created fireball is the slope of the transverse-mass
spectra ($m_{t}=\sqrt{M_{\text{ee}}^{2} + p_{t}^{2}}$). The effective
slope parameter $T_{\mathrm{eff}}$ can be extracted using the fit
function \cite{Renk:2006qr}
\begin{equation} \frac{\dd N}{m_{t} \, \dd m_{t}} = \mathrm{C} \cdot
\exp\left(-\frac{m_{t}}{T_{\mathrm{eff}}}\right).
\end{equation} 
Note that $T_{\mathrm{eff}}$ must not be confounded with the temperature
of the medium, as the transverse momentum distribution from a thermal
source is not only determined by the temperature: The radial flow of the
system leads to a significant blue-shift of the $m_t$ spectra as well
\cite{Rapp:2013nxa}. The effective slope parameter for the thermal and
hadronic cocktail (UrQMD) contributions at mid-rapidity for Au+Au and Pb+Pb collisions at three
different energies ($\sqrt{s_{NN}}=19.6$, 200 and 2760\,GeV) is shown in
Figure~\ref{slope}\,(b). The results are presented for different mass bins
ranging from 0.2 to 2.5\,GeV/$c^{2}$. To obtain $T_{\mathrm{eff}}$ the
spectra were fitted in mass-dependent $m_{t}$-ranges corresponding to
$0.5 < p_{t} < 1.8$\,GeV/$c$. Interestingly, the results exhibit
completely different mass-dependencies for thermal and non-thermal
contributions. $T_{\mathrm{eff}}$ of the non-thermal decay contributions reaches a
maximum around the $\rho$ and $\omega$ pole masses (for RHIC energies)
or around 1-1.5\,GeV/$c^{2}$ (at LHC) and falls off when going to lower
or higher energies. Conversely, the slope parameter of the thermal
contributions drops with increasing mass or remains at the same level up
to approximately $1\,\GeV/c^{2}$ and then it shows a strong rise for
higher masses.

The different mass-dependency of $T_{\text{eff}}$ for the thermal and non-thermal decay contributions
can be explained by the different conditions of emission: Where the
thermal source is mainly of hadronic nature, i.e., especially around the
$\rho$ and $\omega$ pole masses, one finds a lower average emission
temperature, compared to the mass ranges dominated by the partonic
contribution. This effect is reflected in the thermal $m_{t}$-slopes. The
increase of $T_{\mathrm{eff}}$ for masses above 1\,GeV/$c^{2}$ is due to
the fact that the thermal high mass emission is suppressed at lower temperatures. In
contrast, the non-thermal hadronic decays mostly occur at a late stage of the
fireball evolution, outside the hot and dense region. This leads to
in general lower slope parameters obtained for the non-thermal contributions
compared to the thermal ones. However, note that there is a difference
between the contributions from the long-lived low-mass $\pi^{0}$ and
$\eta$ mesons, for which one finds the lowest $T_{\mathrm{eff}}$, and
especially the very short-lived freeze-out $\rho$ contribution. In spite
of the fact that here the $\rho$ stems only from cells where no thermal
emission is assumed, in its pole mass region one finds a harder slope
than at the $\pi^{0}$ and $\eta$ dominated low-masses. A reason might
also be that these $\rho$ mesons carry additional momentum due to their
rather late and peripheral origin, compared to the other mesons. The
decrease for higher masses above $1\,\GeV/c^{2}$---which are dominated by
the $\phi$ and still some $\rho$---might be explained by the kinematics
of the microscopic decay processes, where high momenta are naturally
suppressed if a particle with higher mass is produced, and the longer
lifetime of the $\phi$ compared to the $\rho$ meson.

The slope parameters for the thermal emission from the coarse-graining approach are similar to those from a fireball parametrization \cite{Rapp:2013nxa} for RHIC energies,
but for the LHC they seem to be somewhat smaller. However, as already mentioned, it is known that the flow effects are underestimated within the
UrQMD model at high collision energies \cite{Petersen:2006vm}, so that
these differences should be mainly due to a less distinct expansion of
the system and not due to differences in the average temperature. The
same conclusion is suggested by the comparison of dilepton spectra with
experimental data, where we saw an underestimation of the yield for
high-$p_{t}$ (see, e.g., Fig.\,\ref{RHICcentrpt}\,(b)).

\section{\label{sec:Outlook}Conclusions \& Outlook} 
In this paper we have presented dilepton spectra for energies available
at collider energies, obtained with an approach using coarse-grained
UrQMD transport simulations to calculate the thermal dilepton
emission. The results for RHIC energies are compared with the
experimental data from the STAR and PHENIX Collaborations and show good
agreement. Furthermore, we could depict that the newest PHENIX results
collected with the HBD upgrade of the detector are now fully consistent
with the STAR measurements and can both be reproduced within the
coarse-graining approach. The excess above the hadronic cocktail in the
region for $0.3 < M_{\text{e}^{+}\text{e}^{-}} < 0.7\,\GeV/c^{2}$ is
hereby explained by thermal emission from a broadened $\rho$ and the
Quark-Gluon Plasma.

For higher masses above the $\phi$ peak our results lie by tendency
somewhat below the experimental data. This can be mainly ascribed to the
missing implementation of the charm emission, which will be the dominant
source for these high masses. However, our results show that also the
partonic emission gives a significant contribution to the overall yield
in this mass region. Furthermore, a comparison of
different EoS indicates that the thermal dilepton spectrum for $M_{\mathrm{e^{+}e^{-}}}>1$\,GeV/$c^{2}$ might enable one to draw conclusions with regard to the QCD phase
structure and the equation of state if the charm contribution can be
reliably subtracted. The present results are consistent with the
open-charm dilepton spectra obtained using a Langevin approach to
simulate the in-medium effects on the invariant-mass spectra in a transport+hydro hybrid model. These
simulations indicate a strong suppression of the open charm contribution
in hot and dense matter compared to the vacuum case, making up only
roughly 50\% of the total high-mass yield \cite{Lang:2013wya}. In consequence, a
study of dilepton emission including the charm contribution in the
coarse-graining approach would be very instructive for the full
understanding of dilepton emission patterns for higher masses and is planned
for future investigations.

While the energy and centrality dependence of the dilepton production
are well reproduced within the model, the transverse-momentum dependence
shows some deviations from the measurement for higher $p_{t}$, whereas
the (dominant) low-momentum production is quite well described. This
effect is probably connected to an underestimation of the collective
flow in the underlying transport model. One should bear in mind that the
model is purely hadronic and that it might therefore not be able to
describe some dynamical effects adequately, which are due to the
emergence of a partonic phase. However, considering the hadronic nature
of the model, the agreement with experimental data as well as the
spectra from fireball parametrizations is surprisingly good. In
consequence, this substantiates the basic idea of the coarse-graining
approach, namely that the only necessary information regarding the
fireball evolution is the distribution of energy and particle densities (or $T$ and $\mu$,
respectively), if one wants to determine the dilepton emission.

Together with the previous results for SIS\,18, FAIR and CERN-SPS
energies, the coarse-graining approach has proven a successful tool for
the theoretical description of dilepton production in heavy-ion
collisions over the whole domain of presently available energies,
corresponding to a range of $\sqrt{s_{NN}}$ which spans over three
orders of magnitude. It is, nevertheless, also apparent that the
coarse-graining approach in its present form can not fully substitute a
study of the QCD phase structure based on a microscopic picture of the
fireball evolution including the effects from the creation of a
deconfined phase of quasifree quarks and gluons. This will be important,
e.g., for the study of the anisotropic flow of electromagnetic probes.

\begin{acknowledgments} 
  The authors especially thank Ralf Rapp for providing the
  parametrizations of the spectral functions. S.~E.~acknowledges Jan
  Steinheimer for valuable and fruitful discussions. This work was
  supported by the Hessian Initiative for Excellence (LOEWE) through the
  Helmholtz International Center for FAIR (HIC for FAIR), the 
  Bundesministerium für Bildung und Forschung, Germany (BMBF) and the 
  Helmholtz-Gemeinschaft throug the Research
  School for Quark-Matter Studies (H-QM). The computational resources
  for this work were provided by the LOEWE-CSC.
\end{acknowledgments}

\bibliography{Bibliothek}

\end{document}